\documentclass[acmsmall, screen]{acmart}
\AtBeginDocument{%
  \providecommand\BibTeX{{%
    \normalfont B\kern-0.5em{\scshape i\kern-0.25em b}\kern-0.8em\TeX}}}

\definecolor{flatgreen}{HTML}{b7f4d8}
\definecolor{flatred}{HTML}{ffcd02}
\definecolor{lightgray}{gray}{0.9}
\definecolor{nolan_green}{HTML}{119F57}
\definecolor{nolan_yellow}{HTML}{F7AB00}
\definecolor{nolan_red}{HTML}{E74C3C}

\usepackage{comment}
\usepackage{listings}
\usepackage{xcolor}
\usepackage{colortbl}
\usepackage{soul}
\usepackage{fontawesome}
\usepackage{array,multirow,graphicx}
\usepackage{float}
\usepackage{pifont}
\usepackage{caption}
\usepackage{balance}
\usepackage[most]{tcolorbox}
\usepackage{tcolorbox}
\usepackage{listings}
\usepackage{amsmath}
\usepackage{framed}
\usepackage{siunitx}
\usepackage{bm}
\usepackage{booktabs}
\usepackage{enumitem}
\usepackage{tipa} %
\usepackage{tikz}
\usepackage{pgf-pie}
\usetikzlibrary{arrows, positioning, shadows}

\usepackage{tikz}
\usepackage[normalem]{ulem}
\usepackage{marginnote}
\usepackage[colorinlistoftodos]{todonotes}

\presetkeys%
    {todonotes}%
    {size=\tiny}
    {}

\definecolor{mGreen}{rgb}{0,0.6,0}
\definecolor{mGray}{rgb}{0.5,0.5,0.5}
\definecolor{mPurple}{rgb}{0.58,0,0.82}
\definecolor{backgroundColour}{rgb}{0.95,0.95,0.92}

\lstdefinestyle{CStyle}{
    backgroundcolor=\color{backgroundColour},
    commentstyle=\color{mGreen},
    keywordstyle=\color{magenta},
    numberstyle=\tiny\color{mGray},
    stringstyle=\color{mPurple},
    basicstyle=\footnotesize,
    breakatwhitespace=false,
    breaklines=true,
    captionpos=b,
    keepspaces=true,
    numbers=left,
    numbersep=5pt,
    showspaces=false,
    showstringspaces=false,
    showtabs=false,
    tabsize=2,
    language=C
}
\usepackage{epigraph}
\usepackage{enumitem}
\usepackage[flushleft]{threeparttable}
\usepackage{multirow}
\usepackage{afterpage}
\usepackage[ruled,linesnumbered]{algorithm2e}
\usepackage{bm}
\usepackage{url}
\usepackage[colorinlistoftodos]{todonotes}
\usepackage{subcaption}
\usepackage{framed}
\SetAlFnt{\small}
\SetAlCapFnt{\small}
\SetAlCapNameFnt{\small}
\usepackage{algorithmic}
\algsetup{linenosize=\small}
\SetKwInput{KwInput}{Input}                
\SetKwInput{KwOutput}{Output}              

\usepackage{titlesec}

\titleformat{\paragraph}
{\normalfont\normalsize\itshape}{\theparagraph}{1em}{}
\titlespacing*{\paragraph}
{0pt}{3.25ex plus 1ex minus .2ex}{1.5ex plus .2ex}

\newboolean{showcomments}
\setboolean{showcomments}{true}
\ifthenelse{\boolean{showcomments}}
{ }

\ifthenelse{\boolean{showcomments}}
{ }

\ifthenelse{\boolean{showcomments}}
{ }

\newcommand{\qs}[2]{\emph{``#1'' (#2)}}
\newcommand{\p}[1]{\emph{(#1)}}
\setcopyright{acmlicensed}
\copyrightyear{2024}
\acmYear{2024}
\acmDOI{XXXXXXX.XXXXXXX}

\begin{document}

\title{AI Tool Use and Adoption in Software Development by Individuals and Organizations: A Grounded Theory Study}

\author{Ze Shi Li}
\email{lize@uvic.ca}
\affiliation{%
  \institution{University of Victoria}
  \country{Victoria, Canada}
}

\author{Nowshin Nawar Arony}
\email{nowshinarony@uvic.ca}
\affiliation{%
  \institution{University of Victoria}
  \country{Victoria, Canada}
}

\author{Ahmed Musa Awon}
\email{ahmedmusa@uvic.ca}
\affiliation{%
  \institution{University of Victoria}
  \country{Victoria, Canada}
}

\author{Daniela Damian}
\email{danielad@uvic.ca}
\affiliation{%
  \institution{University of Victoria}
  \country{Victoria, Canada}
}

\author{Bowen Xu}
\authornote{Bowen Xu is the corresponding author.}
\email{bxu22@ncsu.edu}
\affiliation{%
  \institution{North Carolina State University}
  \country{United States}
}



\begin{abstract}
AI assistance tools such as ChatGPT, Copilot, and Gemini have dramatically impacted the nature of software development in recent years. 
Numerous studies have studied the positive benefits that practitioners have achieved from using these tools in their work. 
While there is a growing body of knowledge regarding the usability aspects of leveraging AI tools, we still lack concrete details on the issues that organizations and practitioners need to consider should they want to explore increasing adoption or use of AI tools.
In this study, we conducted a mixed methods study involving interviews with 26 industry practitioners and 395 survey respondents. 
We found that there are several motives and challenges that impact \emph{individuals} and \emph{organizations} and developed a theory of AI Tool Adoption. 
For example, we found creating a culture of sharing of AI best practices and tips as a key motive for practitioners' adopting and using AI tools.
In total, we identified 2 individual motives, 4 individual challenges, 3 organizational motives, and 3 organizational challenges, and 3 interleaved relationships.
The 3 interleaved relationships act in a push-pull manner 
where motives pull practitioners to increase the use of AI tools and challenges push practitioners away from using AI tools.
\end{abstract}



\received{20 February 2007}
\received[revised]{12 March 2009}
\received[accepted]{5 June 2009}

\maketitle
\section{Introduction} \label{intro}
Large-language models (LLMs), generative artificial intelligence (GAI), and other artificial intelligence (AI) products are quickly transforming the software engineering landscape.
AI assistance tool(s) such as Copilot \cite{github_copilot}, ChatGPT \cite{chatgpt}, and Gemini \cite{gemini} offer the promise of AI powered assistance to help software development teams.
They offer enticing marketing terms such as, ``The AI coding assistant elevating developer workflows." \cite{github_copilot}
Based on early adoption numbers \cite{porter2023chatgpt, microsoft_transcript} and early empirical studies about these tool(s) \cite{liang2024large, peng2023impact, kalliamvakou_research_2022}, these tool(s) are creating substantial value for software practitioners, and their employment organizations who benefit from more happier and productive employees \cite{kalliamvakou_research_2022}.

Early empirical studies have often studied the usability aspects of AI assistance tool(s) \cite{vaithilingam2022expectation, liang2024large}.
Their primary emphasis is on how developers perceive and use AI programming assistants \cite{vaithilingam2022expectation}, as well as usability challenges that developers experience \cite{liang2024large}.
Industry surveys also tried to capture the nature of software developer use, focusing on the specific tasks that developers most commonly rely on AI assistance \cite{StackOverflowSurvey2023, gitlab2023devsecops}.
Another recent study attempted to understand the complexity of generative AI (GAI) adoption in software engineering \cite{russo2023navigating} through survey participants.
In an exploration through surveys, the author looked to understand adoption of generative AI in software engineering through individual, technological, and social levels based on other theoretical models.
The author developed a preliminary model of adoption in software engineering, however, 3/7 of the hypotheses in the model was not supported upon further analysis \cite{russo2023navigating} including impacts from social factors, and personal and environmental factors.
Moreover, the model proposed factors that may influence the adoption but lacked details on how these factors played a role in the adoption. 
Details on the different factors is essential for organizations to understand how they can use the factors to increase adoption of AI tools.
Despite these earlier works, we still lack clarity on what is impacting the adoption and usage of AI tools from practitioners and their organizations, and how can practitioners improve usage?  

To fill this research gap, we conducted a study following socio-technical grounded theory (STGT) \cite{hoda2021socio}, which is an iterative research method that helped us generate our novel theory.
Our research was guided by an initial goal of gathering more understanding regarding: \emph{What impacts AI adoption and use in SE?}
In particular, we conducted 26 interviews with diverse software practitioners from a variety of development roles (e.g., software engineers, developers, DevOps, AI engineers, tech leads, etc.), organizational sizes, and geographic locations.
Additionally, after our basic and advanced stage of theory development, we also conducted a round of theoretical category and relationship validation through a survey with 395 respondents.
Our work improves on Russo's preliminary work by surveying a significantly higher number of participants and developing our theory from semi-structured interviews, offering deeper insights and how organizations can support usage and adoption.
We also developed a theory of 
motives and challenges for adopting and using AI tool(s) from both an individual and organizational level.
For example, practitioners are limited in their work by a lack of training provided by their organization as well as a lack of usage guidelines, which can create uncertainty surrounding privacy risk.
In contrast, we found organizations who provided AI tool training, pulling practitioners to increase usage of AI tools.
We identified 3 push-pull relationships between the motives and challenges that could negatively impact the use and adoption of AI tool(s) if organizations do not adequately handle the challenges.

The contributions of our study are as follows:
\begin{itemize}
    \item \textbf{Identification and Analysis of Motivation:} Our study identified and analyzed 2 individual and 3 organizational motive factors that increase practitioner use and adoption of AI tool(s) in software development.
    \item \textbf{Identification and Analysis of Challenges:} Our study identified and analyzed 4 individual and 3 organizational challenge factors that limit practitioner use and adoption of AI tool(s) in software development.
    \item \textbf{Identification and Analysis of Relationships and Recommendations to Practitioners:} Our study identified and analyzed 9/12 factors that together form 3 different push-pull relationships between motives and challenges. We discuss these relationships in detail and provide recommendations for practitioners, to assist them in navigating the challenges that limit use and adoption of AI tool(s).
    
\end{itemize}
\section{Related Work} \label{related-work}
The recent surge of AI tool(s) such as ChatGPT \cite{chatgpt}, GitHub Copilot \cite{github_copilot}, Llama \cite{llama}, Gemini \cite{gemini}, Claude \cite{claude_ai}, and many more has transformed the software development industry, introducing new capabilities and challenges and sparked discussions about their usage.
This section reviews existing research on AI tool(s) usage and its broader implications on software development practices.

As one of the first AI tool(s) used in the software development industry, GitHub Copilot has been the subject of several research studies.
In the work by Vaithilingam et al. \cite{vaithilingam2022expectation}, the authors studied 24 participants to understand how programmers use and perceive Github Copilot. 
The authors found that while programmers appreciated Copilot for providing a starting point and reducing the need to search online, the tool's generated code was often difficult to debug, understand, and modify. These limitations hindered their task solving effectiveness.
Saki Imai \cite{imai2022github} experimented with 21 participants and found that while GitHub Copilot boosts productivity, it compromises code quality.
Yetistiren et al. \cite{yetistiren2022assessing} assess the quality of generated code provided by GitHub  in terms of validity, correctness, and efficiency.
They found that Copilot successfully generated valid code 91.5\% of the time. However, in terms of correctness, only 28.7\% of code was fully correct, 51.2\% was partially correct, and 20.1\% was incorrect.
The study concluded that GitHub Copilot is promising but it requires further evaluation.
Likewise, Nguyen and Nadi \cite{nguyen2022empirical} and Dakhel et al. \cite{dakhel2023github} evaluated the code generated by Copilot and found that the code quality varies depending on the programming language. 
Furthermore, Barke et al. \cite{barke2023grounded} observed 20 participants using GitHub Copilot and developed a grounded theory that describes two modes in which users interact with the tool: acceleration and exploration. 
In acceleration mode, programmers use Copilot to quickly achieve known goals without disrupting their workflow. 
In exploration mode, programmers use Copilot as a starting points, involving slower, more deliberate interactions with extensive validation. 
Based on their theory, the authors provide design recommendations for future programming assistants.
These studies indicate that while AI tool(s) like GitHub Copilot is promising, there are still challenges 
of code quality and errors which engineers need to validate before using.

Jaber et al. \cite{beganovic2023methods} conducted a systematic literature review of 12 papers on the application of ChatGPT in software development.
The study indicates that ChatGPT has potentials in various domains of software development including automated program repair and bug fixing, programming numerical algorithms, software engineering decision-making and more.
However, the authors indicated concerns about over reliance, reinforcement of misconceptions or biases, and ethical implications of AI use in education and other sectors.
Stemming from ethical concerns of AI tool(s), Pant et al. \cite{pant2024ethics} deployed a survey to understand AI practitioners’ awareness of AI ethics and their challenges in incorporating ethics. 
The study found that most AI practitioners have a good grasp of AI ethics due to their workplace's rules and policies on privacy and security. 
Although formal education and training provide some help, practitioners face challenges related to technological problems and human factors when trying to build ethical AI systems.
Pant et al. \cite{pant2024navigating} further studied the AI/ML fairness and developed a framework outlining the relationship between AI practitioners' understanding of fair AI/ML and their development challenges, the consequences of creating unfair AI/ML, and the strategies used to ensure fairness.
Another area related to AI tool(s) is the trust. 
Johnson et al. \cite{johnson2023make} investigated engineers' trust in software tool(s) by interviewing 18 engineers from within and outside Microsoft. 
The study resulted in the PICSE framework, designed to analyze and enhance trust among engineers in their software tool(s). 
These studies show the potential benefits and challenges of tool(s) like ChatGPT and the current understanding among practitioners.

Prior works have further studied how practitioners leverage AI tool(s) in software development and how these tool(s) impact developer productivity.
Ziegler et al. \cite{ziegler2022productivity} conducted a case study on the impact of GitHub Copilot on developers' perception of productivity by examining how frequently suggestions were accepted by users. 
They identified that developers' perceptions of productivity was mainly influenced by how often they accepted Copilot's suggestions.
Ross et al. \cite{ross2023programmer} developed a Programmer's Assistant and tested with 42 participants if conversational agents based on large language models can help with software development. 
Their study revealed that these agents effectively assist software engineers by generating relevant code and engaging in useful, extended conversations that improve productivity.
In a study by Liang et al. \cite{liang2024large}, the researchers conducted a survey with 410 developers to explore how they use AI programming assistants and the usability challenges they encounter. 
The results highlight that developers primarily use these tool(s) to reduce keystrokes, speed up task completion, and assist with syntax recall.
However, they do not use AI tool(s) for brainstorming solutions. 
The study further unveiled that the AI tool(s) fail to meet specific functional or non-functional requirements and often produced undesired output. 

Other prior works on generative models for programming found that users using natural language for generative models still struggled with challenges because they need to learn the model's ``syntax" \cite{jiang2022discovering}.
A user study where developers tasked to complete a variety of Python programming tasks were not found to achieve statistically significant gains in measurable outcomes using the assistance plugin, in categories such as completion time and correctness \cite{xu2022ide}.

The studies conducted so far have focused on various aspects of these tool(s) such as code generation quality of these tool(s) \cite{yetistiren2022assessing, imai2022github}, impact on developers productivity \cite{ziegler2022productivity}, trust \cite{johnson2023make}, ethics \cite{akbar2023ethical, pant2024ethics}, and usability \cite{vaithilingam2022expectation, liang2024large}.  
More recently, a study by Daniel Russo \cite{russo2023navigating} investigated the adoption of generative AI tool(s) in software development employing surveys based on frameworks like Technology Acceptance Model (TAM), the Diffusion of Innovation Theory (DOI), and the Social Cognitive Theory (SCT). 
The researcher developed an initial model of generative AI tool adoption which included 7 hypothesis.
However, upon further analysis 3 were not supported. 
He concluded that compatibility with existing workflows is the primary driver of AI tool adoption.

Despite previous efforts, there is a lack of understanding of the role that practitioners and their organizations play in adopting and using AI tool(s).
More importantly, given the existing literature that provides evidence about the advantages of leveraging AI tool(s) for software development, for practitioners and organizations to fully reap the benefit of AI tool(s), it is important to understand what should organizations do to increase the use and adoption of AI tool(s).
Thus, our research aims to fill this gap and contribute to improving the adoption and increased use of AI tool(s) in software development.

\begin{figure*}
    \centering
    \includegraphics[width=14cm]{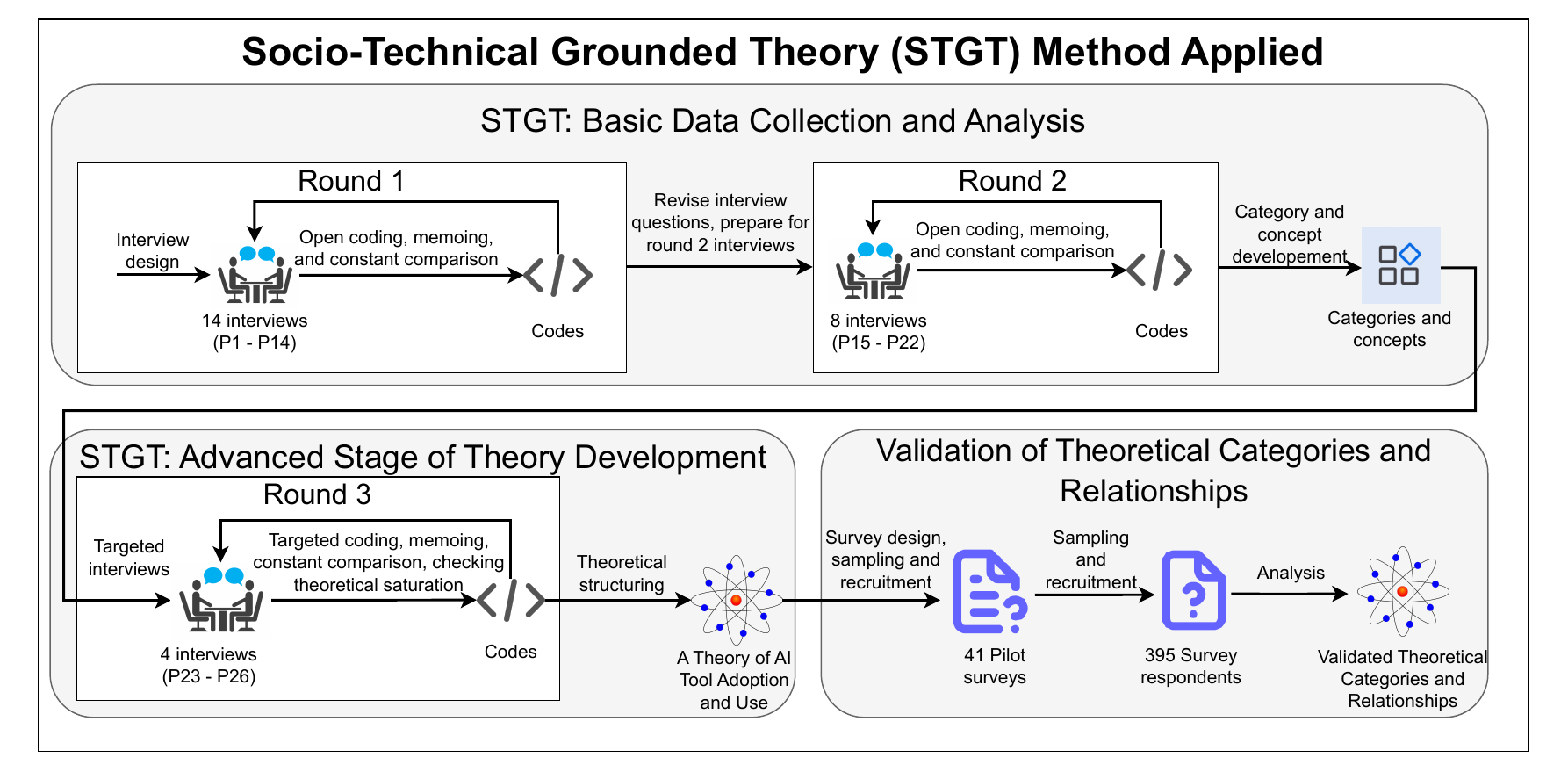}
    \caption{The Methodology of Our Study}
    \label{fig:methodology}
\end{figure*}

\section{Methodology}
Socio-Technical Grounded Theory (STGT) is ideal for studies like ours that involve data collected through interviews, theory development, practice, industry-relevant topics, and the human and social aspects of software engineering \cite{hoda2021socio}.
Since its inception, STGT has been a commonly used research method for software engineering research due to its suitability for technology intensive domains \cite{madampe_emotional_2023, madampe_framework_2023, gunatilake_enablers_2024, graetsch_dealing_2023, gama2023understanding, hidellaarachchi2023understanding}.
Moreover, STGT is best suited for those studies that aim to develop deep understandings through qualitative data \cite{madampe_emotional_2023, madampe_framework_2023,hidellaarachchi2023understanding}. 

STGT is aptly suited for this research topic, since we aim to study the broad emerging area of AI tool usage and adoption by practitioners and reduce challenges in leveraging AI tools in their organizations.
Additionally, the research context of our work closely aligns with the four pillars of the STGT research framework \cite{hoda2021socio} as discussed below.

\begin{itemize}
    \item Socio-Technical \textbf{Phenomenon}: Our study explores the use and adoption of AI tools for practitioners in software organizations. This is a socio-technical phenomenon because usage and adoption of new AI tools may be greatly impacted by multiple factors, such as interactions with tools, technical systems, other team members, organizational structures and processes.
    \item Socio-Technical \textbf{Data, Tools, and Techniques}: Our research team collected interview data via Zoom and Microsoft Teams, and used tools such as Otter.AI\footnote{\url{https://otter.ai/}} to automatically transcribe audio recordings, and Google Sheets to assist with coding and analyzing interviews. 
    \item Socio-Technical \textbf{Domain}: Because of the nature of the software engineering field, there are various socio-technical actors in our study, including but not limited to software developers, AI engineers, software engineers, DevOps engineers, technical leads, and developers. Software engineering teams in organizations and AI assistance tools for software development are socio-technical domains.
    \item Socio-Technical \textbf{Researcher}: The knowledge and skills of the research team encompass a wide breadth of domain and research knowledge and experience. Each interview was conducted by experienced practitioners and/or researchers. 
\end{itemize}

Figure \ref{fig:methodology} illustrates the methodology of our study.
STGT has two variants according to the purpose, e.g., STGT for data analysis \cite{madampe_emotional_2023, madampe_framework_2023, gunatilake_enablers_2024, hidellaarachchi2023understanding} and full STGT study resulting in theory \cite{graetsch_dealing_2023}.
The difference between the two is that a full STGT study results in \emph{``novel, useful, parsimonious, and modifiable, theories grounded in evidence''}, whereas STGT for data analysis applies basic data analysis techniques with other research methods (e.g., case study) and produces preliminary categories and propositions \cite{hoda2021socio}. 
We adhered to the full STGT method prescribed in the STGT guidelines \cite{hoda2021socio} which include two stages \emph{basic data collection and analysis} and \emph{advanced stage of theory development.}
Our application of the full STGT method also involved conducting a questionnaire survey, in a validation stage, to further validate our theoretical categories and relationships \cite{hoda2021socio}.
We conducted 26 interviews in 3 rounds.
Adhering to the STGT \cite{hoda2021socio} process, the basic data collection involved interviews in two rounds: (1) round 1 with P1 to P14, (2) round 2 with P15 to P22. 
The subsequent advanced stage of theory development involved targeted interviews with 4 participants.  

\subsection{STGT: Basic Stage of Data Collection and Analysis}
The basic stage starts with a lightweight review of the existing literature, known as a lean literature review in STGT \cite{hoda2021socio}.
We began with reviewing the existing research on the use of AI tool(s) in SE which we report in the section Introduction (Section \ref{intro}) and Related Works (Section \ref{related-work}).

We chose semi-structured interviews as our primary data collection approach.
The first 14 interviews revealed new insights related to several aspects such as company culture, data privacy, and company guidelines surrounding AI tool(s) usage.
In the subsequent 8 interviews, we narrowed our questions to these aspects of AI tool use and adoption.
Upon analyzing the data from the second round, the categories and relationships became more prominent, highlighting motives and challenges influencing the AI tool use and adoption.
Then we embarked to the next stage, i.e., the Advanced stage.

\subsubsection{Interview Design}
Our initial literature review helped to partially design the interview questions. 
In the literature review, we found survey studies related to the usability of the AI tools and industry oriented surveys. 
For example, the study by Liang et al., \cite{liang2022understanding} surveyed developers on topics such as \emph{why and how often AI tool(s) are used, strategies used to these tools work better, and why developers give up on using the AI-generated code.}
We found another study conducted by Stack Overflow \cite{StackOverflowSurvey2023}, where they surveyed practitioners about their perceptions of AI tool(s) and how these tools may or may not impact their workflows. 
The study revealed that practitioners use AI tool(s) throughout the various phases of Software Development Life Cycle (SDLC) for tasks like coding, debugging, documentation, learning about codebase, code commit and review, testing, as well as deployment and monitoring. 
Thus, we developed our semi-structured interview questions by building on top of these studies and following the general interview guide \cite{patton2002qualitative}, as it is widely used by researchers \cite{alahyari2017study, alami2022journey, itkonen2012role, raghavan2020impact}.

We prepared 40 broad base questions including questions such as 
\emph{1) How do you decide which AI tools to use? 2) How does using AI tools impact work assignments? 3) How do you apply these tools to your workflow?}
We asked \emph{how AI tools impact work assignments}, because previous literature indicated that developers \emph{saved time} from using AI tools. 
Hence, we were interested to see if these time improvements resulted in a difference in how practitioners assigned tasks and divided work.

Benefiting from our careful interview design, we were able to find new insights about potential influences on AI tool use and adoption, such as \textbf{culture of sharing insights with peers}.
For example, from P7's interview we found that they share prompting techniques in their company Slack channel, \qs{We use a company Slack channel, so we have several channels, \textbf{sometimes we just share, how do you prompt} like best way to prompt some research papers? Like React prompting.}{P7}
We also found discussions related to data privacy, peer judgment, lack of guidelines, and more.
These topics emerged as motives and challenges that influenced the AI tool use and adoption in the industry.
The semi-structured nature of our interviews enabled us to unveil these findings, as it allowed us to ask probing questions and delve deeper into the emerging topics. 
Furthermore, STGT methodology emphasizes iterative and incremental research, so we were able to adjust our interview questions in subsequent interviews and ask questions that are typically not supported in survey-based approach.
Hence, we updated our questions to the emerged topics, and added 15 new questions. 
The new set included questions like \emph{``Is there a culture of sharing practices or strategies for using AI tools within your company?''}, as we found many participants highlighting that talking to peers often taught new prompting methods and tips to try on the AI tools. 
A subset of the new questions also included, \emph{1) Does your company provide any guidelines and restrictions on using AI tools? 2) Ever since the new AI tools have come out, how has the company culture changed?} 
The full set of interview questions are included in our replication package.\footnote{\url{https://zenodo.org/records/12165737}}

\subsubsection{Sampling and Recruitment:}
We began participant recruitment by using convenience sampling \cite{baltes2022sampling} and invited industry practitioners from our personal contacts to participate in the semi-structured interviews. 
The initial selection criteria included any software practitioner who uses AI tool(s) for tasks related to software development. 
We expanded our reach through recommendations from previous interviewees. 
To mitigate bias, we aimed to interview practitioners from diverse company sizes, years of experience, and countries to help us understand the overall worldwide usage and adoption of the AI tool(s).  
In total, we interviewed 26 participants from Asia, Europe, North America, Oceania, and South America, whose experience in the software industry varied from 1 to 15 years. 
Their companies also vary in size with small, medium, large, and extra-large-sized.
Table \ref{tab:interviewee_demographics} shows the demographic information of our participants. 

\begin{figure*}
    \centering
    \includegraphics[width=14cm]{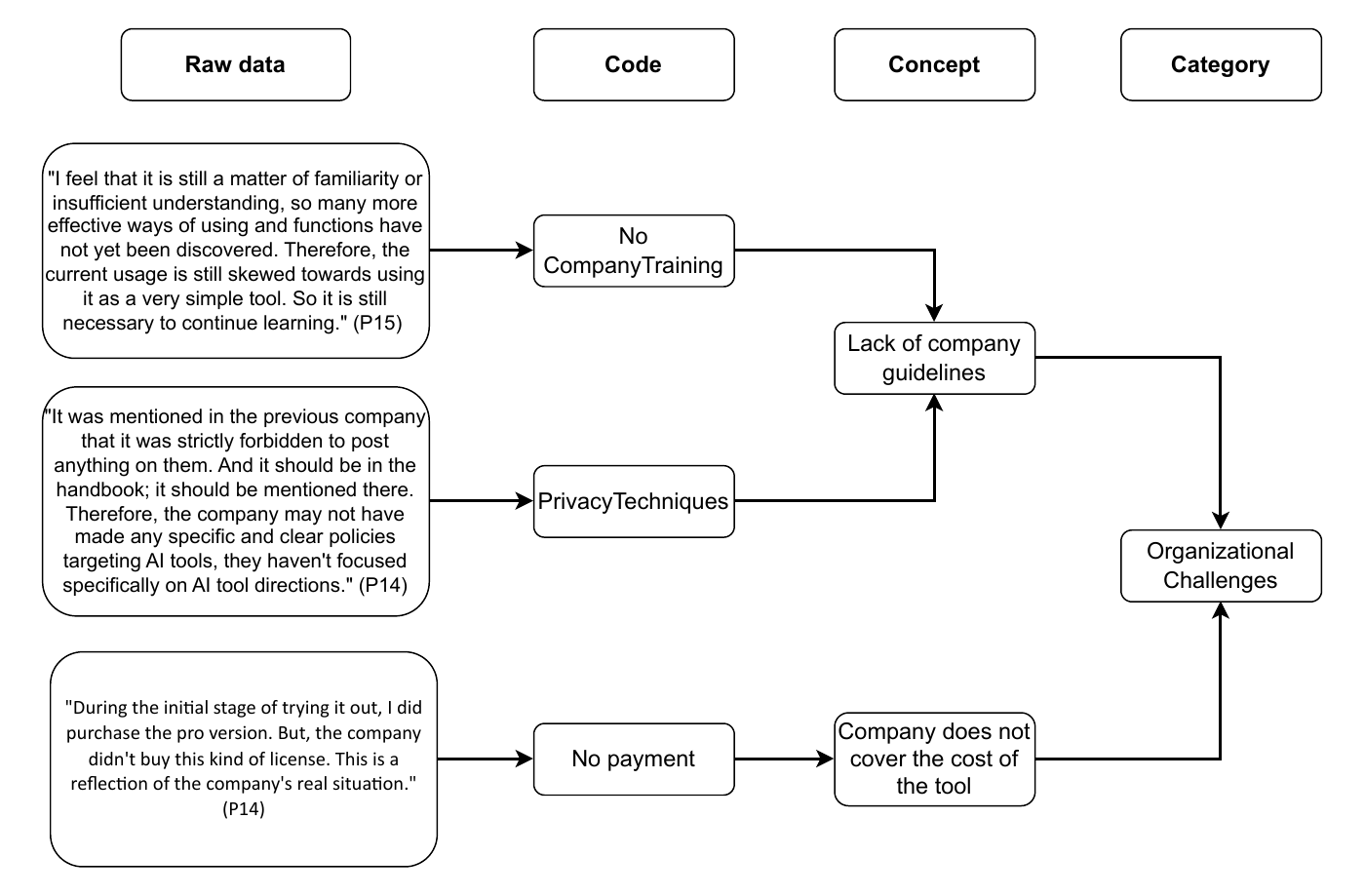}
    \caption{Example Coding of Raw Quotes}
    \label{fig:raw_quote}
\end{figure*}

\subsubsection{Collection and Analysis Procedure:}
In the first round of interviews, we interviewed 14 participants (P1 - P14). 
Each interview was conducted via Zoom and lasted approximately 45-60 minutes. 
At the beginning of each interview, we informed the participants that their interview would be recorded for transcription purposes and would remain confidential.
To analyze the transcripts three authors used open coding to inductively assign codes \cite{strauss1990basics}. 
As new codes emerged we employed a constant comparison method and regularly met to discuss and compare the derived codes.  
We further prepared memos for each interview to reflect on the key points of the interview and draw connections between the codes. 
The codes from the 14 interviews pointed us toward new topics, e.g., discussion with peers, data privacy, peer judgment, lack of guidelines, and more.
Hence, we updated our questions accordingly using these insights and narrowed our focus towards the aspects that influence AI tool use and adoption.

At this stage, we applied theoretical sampling which is the process of narrowing the focus of the data collection to the main findings that emerged from the analysis process. 
The process of theoretical sampling also involves updating the interview questions based on the emerging concepts, categories, and relationships, rather than keeping them broad as previous interviews \cite{hoda2021socio}.
Thus, we conduct round 2 of the interviews which involved 8 participants (P15 - P22) using the updated questions.   
We continued open coding and memoing the interviews. 
From the 22 interview transcripts, we first generated 48 codes. 
Then, we categorized the 48 codes into 12 concepts. 
We refer to these 12 concepts as motives and challenges \emph{factors} that influenced the practitioners' AI tool use and adoption. 
For example, \emph{discussion with peers} motivated a practitioner to use AI tool(s) during their work. 
Whereas, a \emph{lack of guidelines} in the company made it challenging for them to use AI tools, as they were unsure about how much information they could put into the tool.
We show an example of applying STGT analysis from raw quote to category in Figure \ref{fig:raw_quote}.

During the last 3 interviews of round 2, no new insights emerged, and we found detailed and prominent concepts.
Thus, we reached the end of the Basic Stage. 
However, by the end of the Basic Stage of Data Collection and Analysis, we still lacked a full theory. 
Therefore, in the next stage, we proceeded to the Advanced Stage of Theory Development with the ``Emergent Mode'' \cite{hoda2021socio}.
The STGT guidelines \cite{hoda2021socio} defines the Emergent Mode as, 

\epigraph{\textit{``Enabling the emergence of theory through the iterative targeted data collection and analysis step which ends with theoretical saturation and results in a mature theory.''}}{}

\begin{table}[t!]
    \centering
    \small
    \caption{Interviewee Demographics}
     \begin{threeparttable}
        
    \begin{tabular}{p{0.7cm}p{3cm}p{1cm}p{1cm}p{2cm}} \toprule
    P\# & Role & Exp &  Size & Continent \\ \midrule
     P1 & DevOps & 2  &  M   &  North America \\ 
     P2 & Senior Dev. &  6 & L   & Asia \\
     P3 & Developer & 3  &  S  &  Asia \\
     P4 & Senior Dev. & 6  & M   & Asia \\
     P5 & Senior Fullstack & 8  & M   &   North America \\
     P6 & DevOps &  3 &  M  &  North America \\
     P7 & Developer &  3 &  L  &  Europe \\
     P8 & Sw. Engineer & 4  & L   &   North America  \\
     P9 & Sw. Engineer & 2  &  L  &   North America \\
     P10 & Sw. Enginerr  & 5  & S   &  Oceania\\
     P11 & Applied Scientist & 7  & XL   &   North America\\
     P12 & CTO & 5  &  S  &   North America \\
     P13 & AI Engineer & 3  & XL   &  Asia \\
     P14 & Tech Lead &  15 & XL   &  Asia \\
     P15 & Tech Lead &  7 &  XL  & Asia  \\
     P16 & AI Engineer & 7  &  L  &  Asia \\
     P17 & AI Engineer &  10 & XL   & Asia \\
     P18 & Tech Lead & 10  &  XL  &   Asia\\
     P19 & Sw. Architect & 5  &   L &  Europe \\ 
     P20 & Sw. Engineer &  4 &  XL  &  North America \\
     P21 & Sw. Engineer &  5 &   M  & Europe  \\
     P22 & Sw. Engineer  &  3 &   L &  North America \\
     P23 & Sw. Engineer &  3 & L   &  North America \\
     P24 & Sw. Engineer & 6  & M   & South America \\
     P25 & Sw. Engineer &  5 &  L  & South America  \\
     P26 & AI Engineer & 2  &  M  &  North America  \\ \bottomrule
    \end{tabular}
     \begin{tablenotes}
      \tiny
      \item  Organizational size: S (fewer than 50 employees), M (between 50 and 249 employees), L (between 250 and 4999 employees), and XL (5000 employees and more).
    \end{tablenotes}
    \end{threeparttable}
    \label{tab:interviewee_demographics}
\end{table}

\begin{table}[]
    \centering
    \small
    \caption{Number of survey participants who use each tool. Note: We only showed tool(s) that are used by more than 3 participants}
    \begin{tabular}{ll} \toprule
        Tool & Num. Participants (\%) \\ \midrule
        GitHub Copilot & 256 (64.8\%) \\
        ChatGPT & 245 (62.0\%) \\ 
        ChatGPT (Paid)  & 185 (46.8\%) \\
        OpenAI APIs & 137 (34.7\%) \\ 
        Gemini & 124 (31.4\%) \\
        Claude & 91 (23.0\%) \\
        Bing Chat & 85 (21.5\%) \\ 
        Perplexitiy & 75 (19.0\%) \\ 
        Llama 2 & 58 (14.7\%) \\
        Google Bard & 57 (14.4\%) \\
        Tabnine & 49 (12.4\%) \\
        Codeium & 41 (10.4\%) \\ 
        Poe & 27 (6.8\%) \\ 
        Amazon CodeWhisperer & 25 (6.3\%) \\ 
        Mistral & 6 (1.5\%) \\
        Phind & 4 (1.0\%) \\
        \bottomrule 
    \end{tabular}

    \label{tab:tool_usage_stats}
\end{table}

\subsection{STGT: Advanced Stage of Theory Development}

We conducted two steps in this advanced stage: 1) \emph{targeted data collection and analysis}, and 2) \emph{theoretical structuring}.
Targeted data collection and analysis involves first targeting data sources (i.e., interviewees) to help strengthen our concepts.
The analysis involves targeted coding and constant comparison, which refers to coding the most important concepts from the Basic Stage.


\subsubsection{Targeted Data Collection and Analysis:}
To implement targeted data collection, we conducted a 3rd round of interview with 4 participants (P23 - P26).
In round 3 of the interviews, we focused on strengthening the identified motives and challenges influencing AI tool use and adoption.
We interviewed 4 practitioners, 3 software engineers and 1 AI engineer.
These interviews lasted 30-40 minutes, slightly shorter than the previous round, as the questions at this stage were more targeted and focused.

\subsubsection{Theoretical Structuring:}
The last 4 interviews strengthened the 12 concepts, indicating that we reached theoretical saturation \cite{hoda2021socio}. 
To structure the concepts, we decided to leverage visualization techniques using diagramming software.
From the visualization, we found that the 12 concepts generally fall into four categories: individual motives, individual challenges, organizational motives, and organizational challenges.
We further found a push-pull relationship among the concepts and categories which we denote using double sided arrows. 
We present the concepts, categories, and the push-pull relationship in Figure \ref{fig:theory_figure}.
The 4 motives and challenges and relationships together influence the use and adoption of AI tool(s) which we present as our ``A Theory of AI Tool Use and Adoption for Software Development''. 
In section \ref{sec: theory of adoption}, we provide an in-depth explanation of the theory with example quotes from our interviewees.

\begin{figure}[h!]
\centering
\begin{tikzpicture}
\pie[radius=2]{12.4/1, 15.2/2, 20.3/3, 21.3/4, 13.7/5, 5.8/6, 11.3/7 or more}
\end{tikzpicture}
\caption{Number of AI tools used by each survey participant}
\end{figure}
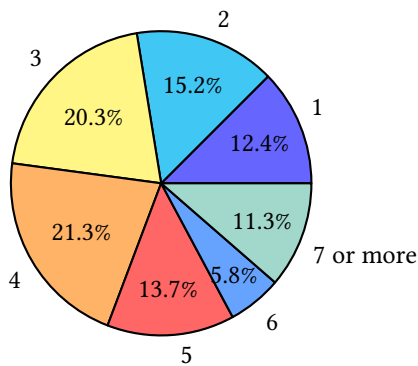

\subsection{Validation of Theoretical Categories and Relationships}
As an additional step, we deployed a survey to validate the categories and relationships developed through the STGT process. 

\subsubsection{Survey Design:}
The survey began with a description of the purpose of our data collection and a set of questions on demographics such as years of experience and number of employees in respondent organization. 
To minimize participant fatigue, we designed a 10 minute survey with few open-answer questions and made all of them optional.
For example, \emph{in your opinion, what would improve your organization's adoption of AI tool(s)?}
For the rest of the questions regarding the factors impacting AI tool use and adoption, we refined the questions so that respondents could easily answer \textit{yes} or \textit{no} or \textit{not applicable} whether a factor impacted their use and adoption. 
For example, \emph{``After I used AI tool(s), my use of tool(s) such as Google and Stack Overflow to help debug or learn has decreased''}.
Furthermore, we asked questions about the relationships between different categories pulling and pushing AI tool use and adoption.
We asked survey questions about every motive and challenge and piloted the survey with 41 participants (i.e., 27 from LinkedIn, 4 from X, 10 from Facebook).
These 41 pilot participants were recruited from open calls to participate from the author(s) social media posts on the three platforms.
It was from the piloting that we found that the survey could be further optimized, therefore, we switched the answers for factors for adoption into \emph{yes} or \textit{no} or \textit{not applicable} as opposed to the standard 5-level Likert scale which we kept for the questions about relationships between different factors for adoption.
The results from the pilot surveys were not included in this study.

\subsubsection{Sampling Strategy:}
To recruit potentially suitable participants, we focused our attention towards GitHub as it provides a platform that allows to identify software professionals who may be using AI assistance tool(s) based on their interests.
We adopted an approach similar to previous studies that used GitHub to recruit survey participants \cite{liang2024large, liang2022understanding, huang2021leaving}.
Similar to Liang et al. \cite{liang2024large}, we identified developers interested in AI assistant programming tool(s), which include \textit{Copilot.Vim \cite{github_copilotvim}, ChatGPT \cite{github_chatgpt}, BetterChatGPT \cite{github_betterchatgpt}, TabNine \cite{github_tabnine}, awesome-chatgpt-prompts \cite{github_awesome_chatgpt_prompts},} and \emph{awesome-ChatGPT-repositories \cite{github_awesome_chatgpt_repo}}.
Using GitHub's API \cite{github_rest}, we collected users who starred any of the aforementioned repositories.
Since it was possible for a user to star multiple repositories, we eliminated duplicate users and also removed users who did not disclose a publicly available email.
Upon filtering we were left with over 20K GitHub users with their respective emails. 
In line with previous survey invitations \cite{liang2024large}, we randomly selected 14,525 users and sent them the survey with a recruitment message. 
As part of our recruitment, we also offered participants (i.e. both pilot and actual survey) to enter in a draw to win one of four \$40 gift cards.
If a participant wanted to opt in for the draw they could fill in their email.
For the full list of survey questions, we provide it in our replication package.\footnote{\url{https://zenodo.org/records/12165737}}

\subsubsection{Survey Analysis:} 
The total number of respondents was 395, resulting in a response rate of 2.7\% (395/14525).
For close-ended questions, we counted the number of responses for each question.
In contrast, for open-ended questions, two co-authors conducted open coding \cite{saldana2021coding}, memoing, and constant comparison. 
The number of responses for each question differed as participants were allowed to answer ``not applicable" or leave the survey at any time. 
Our response rate is in line with other similar surveys \cite{liang2022understanding, liang2024large}.
We note that our response rate was initially around 3.2\% for the first 301 responses (9,406).
However, the response rate dropped for the final batch (1.8\%, 94/5,119) because instead of sending emails directly from our own emails, we had to rely on SurveyMonkey's built-in email service, which resulted in many unopened emails, potentially going straight to recipients' junk inbox.
Our respondents are from 58 different countries and from the continents of Africa (10 people), Asia (159 people), Europe (101 people), Oceania (10 people), South America (25 people), North America (87 people), and 3 undeclared. 
In addition, our participants exhibit a diverse professional software-related experience with $<$ 1 year of experience (30 people), 1 $-$ 3 years of experience (78 people), 3-5 years (79 people), 5-10 years (113 people), 10+ years of experience (95 people).
Finally, our participants are also working in varied sizes of organizations, $< 50$ people organizations (203 people), $51-249$ people organizations (65 people), $250-4999$ people organizations (65 people), $> 5000$ people organizations (62 people).
This breakdown of the employer sizes fits the economic reality.
For countries in the OECD, small and medium enterprises represent 99\% of all firms \cite{matt2020sme}.  
In Table \ref{tab:tool_usage_stats}, we provide the AI assistance tool(s) that are used by our survey participants.
As shown in the table, there are numerous tool(s) that software professionals use, but the most prolific ones in our study are Copilot, ChatGPT (Paid) ChatGPT (free), OpenAI APIs, Gemini, and Claude.
There are some less popular tool(s) that were mentioned only once or twice in total across the 301 participants, so we exclude them in the table.

\begin{figure*}
   
    \includegraphics[width=15cm]{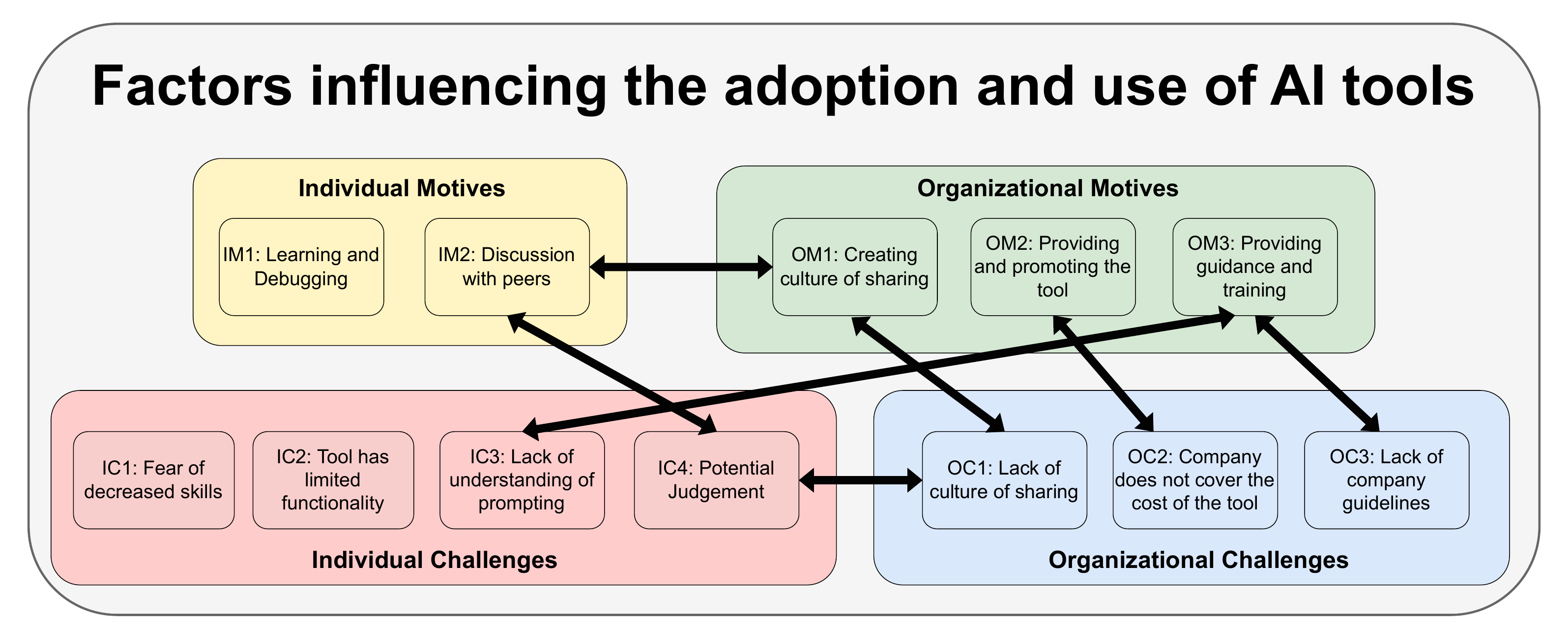}
    \caption{A Theory of AI Tool Use and Adoption in Software Engineering}
    \label{fig:theory_figure}
    \small
    We use double sided arrows to make connections between concepts to show there is a relationship between the two or more concepts.
\end{figure*}

\section{A Theory of AI Tool Use and Adoption for Software Development}\label{sec: theory of adoption}
This section details our theory (Figure \ref{fig:theory_figure}) on the individual and organizational motives and challenges that impact the use and adoption AI tool(s). 
The theory encompasses the motives (i.e., pull) and challenges (i.e., push).
In Table \ref{tab:motives_tab1} and Table \ref{tab:limitations_tab1}, we also detail the number of survey respondents who experienced and/or agree with each factor.
For Tables \ref{tab:motives_tab1} and \ref{tab:limitations_tab1}, we also provide finer grained details for example \emph{IM1 Learning and debugging} contain additional details: IM1.1 learning about new code, programming languages, or concepts easier, IM1.2 debugging errors or problems easier, IM1.3 searching new code, programming languages or concepts easier.

\begin{table}[!htb]
\centering
\small
\caption{Survey responses for motives that increase AI tool use and adoption}
\begin{tabular}{lp{10cm}l}
\toprule
ID & Motives & \#Votes (Ratios) \\ \midrule
\rowcolor[gray]{.90}\multicolumn{3}{c}{\textbf{Individual Motives}} \\ \midrule
\textbf{IM1} & \textbf{Learning and debugging} & \\\midrule
IM1.1          & I realized AI tool(s) make learning about new code, programming languages, or concepts easier, my AI tool(s) usage has increased  & 265/290 (91.4\%) \\
IM1.2          & I realized AI tool(s) make debugging my errors or problems easier & 242/291 (83.2\%) \\
IM1.3        & I realized AI tool(s) make searching about new code, programming languages, or concepts easier	 & 239/290 (82.4\%) \\
IM1.4   &   I realized AI tool(s) reduce my use of tool(s) such as Google and Stack Overflow to help debug or learn & 200/295 (67.8\%) \\ \midrule


\textbf{IM2} & \textbf{Discussion with peers} \\ \midrule
IM2.1          & We started having regular discussions about our use of AI tool(s) (e.g., best practices for prompting in AI tool(s)) within the team and/or the company &  155/247 (62.8\%) \\
IM2.2    &  The employee wants to improve their AI usage skill by sharing  &  187/224 (83.5\%) \\
\midrule
\rowcolor[gray]{.90}\multicolumn{3}{c}{\textbf{Organizational Motivates}} \\ \midrule
\textbf{OM1} & \textbf{Creating culture of sharing} \\ \midrule
OM1.1 & As company created a space for employees to openly discuss about the use of AI tool(s) (e.g., events, slack channels) & 128/189 (67.8\%) \\ 
OM1.2   & The employee's team has a strong culture of sharing insights & 188/223  (84.3\%) \\
OM1.3    &  The employee's company has a strong culture of sharing insights  &  151/213  (70.9\%) \\
\midrule

\textbf{OM2}  & \textbf{Providing and promoting the tool} &  \\ \midrule
OM2.1 & As company paid for the use of AI tool (e.g., provided subscription) & 124/191 (64.9\%) \\
OM2.2 & Because company provides AI tool plugin that is powered by external APIs (e.g., GPT4) & 100/166 (60.2\%) \\
OM2.3 & Company started notifying us on the availability of AI tool(s) (e.g., announced in a meeting) & 106/177 (59.9\%)  \\
OM2.4 & Company started motivating us on the use of AI tool(s) (i.e., sending us email reminders) & 93/168 (55.4\%) \\
OM2.5 & Company mandated us to update our manager about our use of AI tool(s)	 &  61/139 (43.9\%) \\

\midrule
\textbf{OM3}  & \textbf{Providing guidance and training} &  \\ \midrule
OM3.1 & Once company started providing training on prompting AI tool(s) & 79/123 (64.2\%) \\ 
OM3.2 &  After company provided a policy with rules and restrictions of AI tool(s) usage & 82/145 (56.6\%) \\

OM3.3 & Company provides an integration/tool that helps sanitize my prompt to protect company specific data	 &  74/122 (60.7\%) \\ 

\bottomrule
\end{tabular}
\label{tab:motives_tab1}
\end{table}

\begin{table}[!htb]
\small
\centering
\caption{Survey responses for challenges that limited AI tool use and adoption}
\begin{tabular}{p{0.6cm}p{10.8cm}l}
\toprule
ID & Challenges & \#Votes (Ratios) \\ \midrule
\rowcolor[gray]{.90}\multicolumn{3}{c}{\textbf{Individual Challenges}} \\ \midrule
\textbf{IC1}  & \textbf{Fear of decreased skills}  &  \\ \midrule
IC1.1   & A fear of a decreased coding ability  & 53/74 (71.6\%) \\
IC1.2          &  A fear of losing learning ability &  39/74  (52.7\%) \\
IC1.3         &  A fear of over relying on AI tool(s) & 56/75 (74.7\%) \\ 
 \midrule
\textbf{IC2}  & \textbf{Limited AI Capabilities for Software Development} &  \\ \midrule
IC2.1   & Tool does not have operation environment information (e.g., versioning, hardware)	  & 90/141 (63.8\%) \\
IC2.2          &  Tool does not have access to codebase and is unable to generate the correct results	 & 125/163  (76.7\%) \\
IC2.3         &  AI tool often refuses to provide a response for my question (e.g., "As a language model, I cannot answer your question") & 84/155 (53.8\%) \\ 

\midrule

\textbf{IC3} & \textbf{Lack of prompting skill} \\ \midrule
IC3.1   &  Due to AI tool(s) as it needs a lot of experimentation before getting a desired result & 146/264 (55.3\%) \\
IC3.2        &  Tool getting me into an infinite loop of prompting and diverging from the actual task & 42/71 (59.2\%) \\
\midrule
\textbf{IC4}  & \textbf{Potential judgment} &  \\ \midrule
IC4.1          &  	Due to potential judgement from peers & 45/218  (20.6\%) \\ \midrule
\rowcolor[gray]{.90}\multicolumn{3}{c}{\textbf{Organizational Challenges}} \\ \midrule
\textbf{OC1}  & \textbf{Lack of culture of sharing} &  \\ \midrule
OC1.1   & 	Everyone does different work, do not have similar project context  & 24/31 (77.4\%) \\
OC1.2   & 	There is no culture of sharing within my team and/or company  & 20/29 (68.9\%) \\
OC1.3   & 	Fear of potential judgment from others when sharing  & 16/30 (53.3\%) \\ \midrule
\textbf{OC2}  & \textbf{Company does not cover the cost of the tool} &  \\  \midrule
OC2.1          &  Due the high prices and you have to pay for it	 & 120/201  (59.7\%) \\
OC2.2         &  Number of API calls allowed in a time interval (e.g., 40 prompts/4 hours on ChatGPT)	 & 73/142 (51.4\%) \\ \midrule
 
\textbf{OC3}  & \textbf{Lack of company guideline} &  \\ \midrule
OC3.1          &  Due to company not providing any usage guideline	 &  68/201 (33.8\%) \\
OC3.2        &  Due to company not providing training	 &  49/197 (24.9\%) \\ 
OC3.3        &  Due to concerns about accidentally leaking company data &  93/201 (46.2\%) \\


\bottomrule
\end{tabular}
\label{tab:limitations_tab1}
\end{table}



\subsection{Individual Motives}
\subsubsection{IM1: Learning and Debugging:}
AI tools improving practitioners' ability to learn and debug was widely reported by our interviewees and surveyees.
The ostensible reason for AI tool use was that practitioners benefit largely from the \emph{time} saved when learning about new code, debugging errors, or searching about concepts.
Prior to AI tools, when writing new code, practitioners relied on a combination of personal knowledge, Stack Overflow, documentation, and Googling to reach their conclusion \cite{wu2019developers}.
However, our participants reported that AI tools such as Copilot and ChatGPT lowered their cost of \emph{time} to generate code, particularly boilerplate or repetitive code \cite{liang2024large}.
A previous survey study also briefly highlighted that practitioners use AI tools for learning, searching and debugging \cite{russo2023navigating}.
However, the study failed to discuss \emph{how the use} of AI tools impact these tasks.
In contrast, our study detailed that there are noticeable differences between time saved from using AI tools between different roles, and that leveraging AI tools for various tasks led to declined use of traditional software searching and debugging tools such as Stack Overflow and Google.

Through using AI tools, almost all of our interviewees reported perceived \emph{time} savings in their work between 5\% to 50\%.
Admittedly, since virtually none of interviewees adopted a systematic metric to measure their work allocation and productivity, we relied on their self reported time savings. 
To our surprise, software developers reported low time saving with a range between 5\%-25\%.
In contrast, people in more AI or research roles such as \emph{P11, P17, P26} reported saving upwards of 50\% in their code related work.
Our initial assumption was that since there were a plethora of training data on backend and frontend code that software engineer's work would be significantly simplified, but due to restrictions such as privacy, which we cover later, the time savings is less than we assumed. 
On the other hand, AI engineers reported to us that their coding work involved a fair amount of structured code regarding models that they are building towards so much of that can be generated with the help of AI tools.

Specifically, 13 interviewees \p{P1, P2, P3, P4, P5, P10, P12, P15, P16, P20, P21, P22, P24} admitted to using AI tool(s) to help learn a new or existing piece of code. 
265/290 (91.4\%) of survey respondents agreed that their use of AI tools increased after determining that AI tool(s) make learning about new code and programming concepts easier.
P12 recollects how they dealt with an assortment of unfamiliar technologies such as dealing with cloud infrastructure while wanting to release a prototype as soon as possible.
\qs{What's the first step to creating this service? I don't know how, so I ask ChatGPT, and it tells me I could use FastAPI. ``But how do I get it online?" Then I asked, ``how to do that", it says to use EC2, so I just hammered away at it, right? And then I got it up and running.
}{P12}
Before the introduction of AI tools, P12 had to expend considerable resources reading and studying manuals regarding these technologies.



Additionally, our interviews highlighted the advantages from searching with AI about unfamiliar problems.
Prior to AI tools, a developer would probably ask Stack Overflow or search Google for a few possible solutions \cite{wu2019developers, ponzanelli2014mining, yang2016query}, but \emph{identifying the most relevant search terms} may be difficult. 
\qs{I asked ChatGPT and it gives me some possible ways. Then I just pick those suggestions and do some searching on Google and then read some blogs. 
}{P2}
To this point, 239/290 (81.4\$) survey respondents agreed that AI tool(s) make searching about new code and programming languages easier, this is in line with 2023 Stack Overflow survey \cite{StackOverflowSurvey2023}.

Similarly, relying on AI tools to debug code errors was widely reported by our interviewees \p{P1, P3, P5, P6, P9, P11-P15, P17, P20, P21, P23, P25, P26} and surveyees (242/291 (83.2\%)).
Before the use of AI tools, practitioners Googled stack trace or other messages to determine potential causes for issues \cite{li2022debugging, xia2017developers, sadowski2015developers}. 
With the ability of AI tools that allowed uploading screenshots, interviewees explained that sifting through Stack Overflow pages is rather unnecessary. 
\qs{The advantage is that I don't have to spend time sifting through Stack Overflow to see which post actually resembles my problem. Often on Stack Overflow, even though many problems seem similar, they are not exactly the same.}{P19}

The reduction in effort has led to our interviewees and surveyees (200/295 (67.8\%)) to describe that their reliance on help from Stack Overflow and Google for debugging has decreased. 
\qs{
Our internal engineers now, whether they are working on algorithms or on research and development, use [AI tools] to debug, and they no longer need to go to Google anymore using GPT4 for debugging.}{P17}.

\subsubsection{IM2: Discussion with peers:}
Practitioners in our study described that a key motive for increasing use of AI tools is taking on a personal initiative in sharing experiences and successes with AI tools with others in their organization.  
16 interviewees mentioned having experienced discussions with peers about using AI tool(s) and 155/247 (62.8\%) of surveyees agreed that their AI tool usage increased after having regular discussions about their use of tool(s) within the team and/or the organization.

One reason of discussion with others in the team and company was beneficial is that despite the seemingly high popularity of AI tools, as reported by our participants there are still many colleagues who are not yet aware nor tried using them.
As described by Geoffrey Moore \cite{moore1999crossing}, early innovators and early adopters of emerging technologies and products represent only a segment of the overall population.
Hence, our participants recounted examples of colleagues who emerged as strong proponents of AI tools as a result of discussions with colleagues.
\qs{We were coordinating with a third-party vendor to adjust some logs and monitoring. A teammate from our group could not figure it out even after a day. \textbf{So I told him to try ChatGPT and it helped him write the function.} 
\textbf{He shared about [this incident] in our big group chat, saying, ``Look at this, it's amazing, I was able to write such a function."}}{P15}.
The colleague was aware of ChatGPT, but never actually attempted it for their work and now they share AI tool's usefulness to other colleagues.
It exemplifies the organic spread of AI tool use and adoption driven by positive user experience and peer recommendations.

Another reason that discussion is beneficial is that it facilitated a debate of ideas and insights, and practitioners could improve their own skills from this discussion. 
187/224 (83.5\%) survey participants reported that their AI usage increased from sharing best practices and improving their own usage skills.
These discussion sessions allowed practitioners to \emph{learn} about how others in the organization were using AI tools and tips to achieve the best prompts.
\qs{We've done a few presentations ourselves everybody has tried and \textbf{we wanted to present what we have tried doing with AI} ... I think everybody's really curious. They want to try different things.}{P23}

Our interviewees also told us about how tips are shared and others within the team benefit from the collective knowledge base. 
\qs{We talk about it during team meetings to figure out what has or has not been working for us. We do talk about what kind of tool(s) we are using, what kind of processes need to be we need to be using to help us do our work better. 
}{P9} 
These meetings also facilitate discussions about tips such as  \qs{ideas about the best way to prompt ChatGPT to get the kind of information you want.}{P20}


From our survey responses, we also found a difference between discussing and sharing with peers. 
Respondents from Asia had the highest agreement rate for both IM2.1 and IM2.2 at 90.6\% (77/85) and 78\% (78/100).
In contrast, other continents with large respondent populations like North America and Europe had agreements of IM2.1 82.1\% (46/56), IM2.2 54.5\% (30/55) and IM2.1 78.2\% (43/55), IM2.2 47.5\% (29/61).
One probable reason for this phenomenon could be the greater emphasis on collectivist benefits found in many Asian cultures \cite{michailova2006national}.
The opposite is often observed in western cultures where individualism is given higher priority \cite{michailova2006national}.
The sentiment of everybody needing to become acquainted with these tools is shared by P16, \qs{It's best to start using and learning or mastering this tool as soon as possible because I believe that in the future, it will become as essential as tools like Excel are today, something that everybody values and everybody needs to use}{P16}.

\subsection{Individual Challenges}
\subsubsection{IC1: Fear of decreased skills:} \label{ic1}
A variety of fears related to the use of AI tool(s) were articulated by participants, including 9 interviewees \p{P1, P2, P4, P11, P12,  P14,  P17, P19, and P20}. 
Although AI tools assist in boosting learning and debugging, which lead to perceived speed ups in time, participants also reported fears that hinder the use and adoption of AI tools.
53/74 (71.6\%) and 39/74 (52.7\%) of survey respondents agreed that their usage of AI tool(s) is challenged by their fears for a decreased coding ability and losing learning ability.

Despite the ability for AI tools to generate working code, P2 and P1 stresses the importance of understanding the generated code
\qs{If you're copying code from the ChatGPT, then you have to understand and you have to think about it.
\textbf{We are using ChatGPT or other tool(s) for increasing our productivity, but not damaging our thinking, or our creativity}.}{P2}
\qs{\textbf{I don't really like the idea of going to chatgpt, copying a bunch of code that somebody else implemented in a repository.} Even even with stuff prompts that I am entering, I try to think twice.}{P1}

Although AI tools offer access to rapid generation of potential solutions to various software development challenges, our interviewees described that over-reliance could lead to a knowledge gap, affecting the developers' ability to grow and adapt. 
Over-reliance may hinder junior developers skills,  whereas prior to AI tools, juniors could grow by learning and understanding the codebase.
\qs{If they are juniors, if they can explore the codebase, they will learn more. ...
\textbf{there will be some knowledge gap about codebase for a developer who did not use ChatGPT and a developer who, who is using ChatGPT from the beginning of his software engineering career}.}{P4}
This worry was exhibited in how surveyees responded to our questions regarding whether they have a fear of decreased skills IC1.1-IC1.4.
For surveyees that had over 10+ years of professional experience, that agreed with IC1.1) 88.9\% (8/9), IC1.2) 88.9\% (8/9), IC1.3) 87.5\% (7/8), and IC1.4) 100\% (6/6).
Contrastingly, respondents with $<$ 1 year of professional experience agreed with IC1.1) 60.0\% (6/10), IC1.2) 40\% (4/10), IC1.3) 60\% (6/10), and IC1.4) 44.4\% (4/9).

On a personal level our participants described the risk of decreased creativity and critical thinking as a result of over-reliance. 
\qs{\textbf{When we use ChatGPT people don't want to think for themselves anymore at all}... 
\textbf{Because over time, if you're not engaging your mind, you're engaging yourself in thinking about things, you become less creative, and you become more dependable on these tool(s).}}{P20}



\subsubsection{IC2: Limited AI Capabilities for Software Development:}
One of the major challenges of AI tools that impedes use and adoption is their limited capabilities in specific instances, as mentioned by our interviewees. 
Nearly all participants aspired for more functionalities, but 14 of them \p{P1, P5, P9-P12, P14, P17, P20-P22, P23, P25, P26} explicitly desired additional improvements such as access to the operational environment and codebase. 

Our interviewees highlighted that the lack of awareness of the operational environment, including specific software versions, hardware configurations, or the nuances of the programming languages in use often becomes a bottleneck.
This leads to AI generated responses that may be incompatible with the developer's actual working environment. 
P11 explains this issue, 
\qs{Current code can only understand the code, but it cannot comprehend the underlying running environment. I think this is the biggest problem. 
If it doesn't understand the running environment, it can only see the most surface-level issues. 
But if it can see what the \textbf{environment is like}, I think it could go to another level.}{P11}
Prior to AI tools, developers would have to dig through Google or Stack Overflow and hope to find an answer with a compatible environment.
It is easier now for AI to generate a ``correct" sounding answer, but without knowing the operational environment, developers run into the same problem of hoping to find the right solution for their given circumstance.  
As P1 described, \textit{[AI response] is just sort of like a quality distillation of averages of what other people have already agreed upon in some sort of a public context.}
90/141 (63.8\%) of surveyees also felt that tools not having operation environment information was a challenge that reduced their AI use.

Our participants also described that tool(s) most times do not have access to a user's codebase.
93/109 (85.3\%) survey participants agreed that their usage of AI tool(s) was limited by tooling not having access to the codebase and inability to generate the correct results.
This limitation hampers the ability to generate context-aware solutions or debug effectively. 
Without direct access to the codebase, these tool(s) often resort to generic responses that may not be useful.
\qs{It doesn't have the amount domain knowledge that my team member would have. [It] just has general knowledge that the internet has.}{P9}
The limitation also translates into debugging due to a lack of context.
\qs{It's just really unreliable, you usually have to give it the full context. It is a lot of context you [have to] give it or it just doesn't know what to do.}{P8}
To workaround this limitation requires detailing the codebase and context, but this may also lead to concerns about privacy.
As we will describe in Subsection \ref{oc3}, data privacy could have significant revenue and legal ramifications.

The impact of limited AI capabilities is even more noticeable for software organizations working in less common domains or programming languages.
\qs{As you get more and more into specific problems in your domain, it's less helpful.}{P23}
\qs{We use a test framework called [removed], and it's a bit older, and very unintuitive and unfriendly...
But [AI tools] hasn't really been that successful. I wonder if it's because [removed] is not that popular and it didn't really train on that data a lot.}{P5}
For our interviewees who use less popular programming languages, they all indicated that their desired improvement is adding language support for their domain.


Our interviewees also expressed the need to be considerate of using AI tools for writing reports or communicating with clients and other stakeholders.
Tools like ChatGPT are known for its ability to write speeches \cite{herbold2023large}, articles \cite{wu2023brief}, and even research papers \cite{lingard2023writing}, but some interviewees expressed dissatisfaction regarding the superfluous language and unnatural sounding tone from AI tools.
\qs{I don't like its writing style. It's a bit too verbose, and it is very much not my style. 
}{P5}
\qs{The boss can tell at a glance that most of it is written by GPT... 
He doesn't encourage it because he thinks you are genuinely slacking off and you entrusted everything to GPT without full consideration.}{P17}


\subsubsection{IC3: Lack of prompting skill:}
Another notable challenge identified by 9 participants \p{P1, P3, P5, P6, P14-P16, P19, P21} revolves around the difficulty in crafting the right prompts.
The process of finding the right prompt often involves trial and error which can be time consuming. 
146/264 (55.3\%) survey participants agreed that their usage of AI tools is challenged due to AI tool(s) needing a lot of experimentation before getting a desired result.

As P21 described in their case,
\qs{\textbf{I've still been struggling with what kind of prompts to use, how to do it more efficiently}... I feel like for different tasks, it's better to do different prompts.}{P21}
Or as P15 describes, they have a lack of familiarity and understanding of using the AI tool(s).
\qs{\textbf{I feel that it is still a matter of familiarity or insufficient understanding}, so many more effective ways of using and functions have not yet been discovered.}{P15}

We found a noticeable difference in survey respondents who have years of professional experience and more junior practitioners regarding the pain of experimenting with a lot prompts.
Survey respondents who have more than 10+ years of experience reported this challenge in 45.2\% (28/62) cases, whereas 87.5\% (21/24) of respondents with less than 1 year of experience reported this challenge. 
The effectiveness of a prompt can depend on the user's experience and familiarity with the domain.
\qs{\textbf{When you have worked for a certain number of years and are in a certain field, I think you have that knowledge on what to write what the prompt should be}. 
If you have been working for a time, then you know what to write and probably what to search. If you know what to search you can figure out roughly the proper prompt. You can give the AI to give you the best result.}{P3}
These results indicated that senior practitioners are more proficient in crafting the correct prompt and asking the relevant questions to reach their desired solution.
Senior practitioners being more proficient is to be expected, however, this difference may be more pronounced in the age of AI since the barrier to finding \emph{solutions} is lowered. 
Previous literature showed that senior engineers often spend time helping junior engineers with inquiries \cite{meyer2019today}.
\qs{The pressure that entry-level programmers may face in the future, which will likely increase employment difficulties, especially for those at the low to middle end of the spectrum. However, high-end programmers should not be affected.}{P17}
One strategy to lower this concern is prioritizing the need for training on prompting.
\qs{Most engineering peers, including myself, feel that we don't clearly understand the underlying workings of the large models. How to use them more efficiently is actually a very important point.}{P15}
We discuss this in more details in Subsection \ref{om3}.


In subsection \ref{ic1}, we described the fear that AI tools work \emph{too well} and developers become dependent.
However, in instances where AI tools fails to work as expected, 42/71 (59.2\%) of surveyees indicated that they face limitations from going into a loop of prompting, leading away from their actual tasks. 
As noted by P3 and P23, \qs{Cases where I found maybe using the tool, leads to someone being stuck in a loop.}{P3}
\qs{Because sometimes you can get stuck in ``this prompt isn't working", or you're in the loop, trying to get some answers out.}{P23}
When this occurs, practitioners are \qs{pushed backed to using Stack Overflow and Google to get help for debugging.}{P14}
Despite the limited capabilities at times, the overall cost to test AI responses is cheap.
\qs{If I ask it a question, you tell me something that's not very accurate, then I immediately test it and find there's a bug, then I throw it back to the tool, and I say, 'Look, this is a bug.' It says 'sorry,' and then it helps me to fix it.}{P12}

\subsubsection{IC4: Potential judgment:}
We found potential judgment for using AI tool as a concern among some of our participants \p{P4, P5, P14, P22, P24, P25}.
45/218 (20.6\%) of survey participants agreed that their adoption of AI tool(s) may be limited due to potential judgment from peers.
This ratio is not as high as some of the other factors, however, 1/5 of survey respondents feeling a potential judgment from fellow colleagues is not an insignificant number as software engineering is a collaborative activity that requires effective teamwork in communication, coordination, and collaboration \cite{barkhi2006study}.

Our interviewees and surveyees indicated that a reliance on AI could be interpreted as a lack of skill, leading to a reluctance to discuss AI tool use openly, even when it might enhance productivity or creativity.
The existence of \emph{imposter syndrome} (i.e., ``person’s experience of feeling like an intellectual phony despite having outstanding academic and professional credentials" \cite{albusays2021diversity}) has been often experienced by women \cite{clance1978imposter} and other minorities \cite{albusays2021diversity}, but can also impact junior developers and other practitioners with less industry experience. 
\qs{\textbf{I almost don't want my coworkers to know.} And not wanting my coworkers to know I use it for weird reasons. Like \textbf{it makes me less of a programmer or something.} 
}{P5}
A fear of judgment exists in junior developers who may worry about being judged by their senior colleagues for not being proficient enough to code without AI assistance. There may be a perception that using AI indicates incompetence or laziness.
Power dynamics plays a negative role in hierarchical environments, where there can be a power imbalance between junior and senior developers. 
\qs{I guess [me and my teammates], we are not seniors yet. Mostly. Just some juniors and me who's kind of intermediate. So we, with us, we don't have that [negative] perception. But with other teams and seniors, they have this kind of negative perception.}{P25}

The negative connotations lead to interviewees reporting that are unwilling to share or admit to using AI tool(s) as they may feel embarrassed.
\qs{It seems like it could really be possible [people are embarrassed to tell others they use AI tool(s)], which is why I've observed the proportion is so low... 
Then there might be a feeling of embarrassment}{P14}
P14 who is a technical lead mentioned,
\qs{I boldly guess that there must be more people [in my team] who are probably using it secretly.}{P14}

We observed this difference in sentiment in the results of our survey as well.
When we asked surveyees about using AI tools and the feeling of judgment from others, 12\% (6/50) and 13\% (7/53) of practitioners who have 10+ and 5-10 years of experience respectively agreed. 
On the other end of the spectrum, 42\% (8/19) and 19\% (9/47) of practitioners with less than 1 year and 1-3 years of experience agreed. 
In particular, those with less than 1 year of professional experience felt a much greater potential for scrutiny that the other groups.
This may have a been a result of cultural norms within certain organizations that prioritize traditional coding methods over AI tools. 

\subsection{Organizational Motives}
\subsubsection{OM1: Creating culture of sharing:} \label{om1}
From our interviews and surveys, we found that core to organizational motives is fostering a culture of sharing of experiences and practices for using AI tool(s).
Participants \p{P6, P7, P9, P10, P15, P18-P21, P23, P26} described how their companies provided open spaces such as Slack channels to exchange knowledge about AI tool use. 
128/189 (67.8\%) of survey participants agreed that their organization created a space for employees to openly discuss the use of AI tool(s).
For example, \qs{we have a Teams channel and we have something called Tech Tips, where people just either talk about tool(s) or how to use the tool(s) and how to use better prompts.}{P21}

While organizational communication channels facilitate the quick sharing of tips, our participants mentioned that a more centralized storage of best practices and knowledge would be more ideal.
\qs{It would be nice if there's like a more centralized repository where people can just share all their tips there.}{P7}
Prior to the use of AI tools, software organizations would often rely on project management and documentation tools such as Confluence and Jira to store important information \cite{froeschl2020best}.
In the age of AI, tips and tricks for using AI tools represent a new potential boost in productivity that could dramatically improve organization efficiency (e.g., P15 colleague's one day workload solved within minutes).
Hence, as P7 alluded, a single, accessible, and searchable knowledge base would provide potential benefits to streamline the process of learning and leveraging AI tool. 

In contrast to organizations where individuals worry about a negative perception for discussing the use of AI tool(s), practitioners described an environment where sharing about AI tool(s) is met with enthusiasm rather than embarrassment.
\qs{Actually, nobody is feeling embarrassed or anything, it's more like everyone is finding this thing quite magical and wondering if it can play a significant role.}{P15}
The narrative is less fearing judgment over using AI tools and more feeling excitement about their potential. 
Our findings indicated that when companies facilitate a culture of sharing, it supports the dissemination of knowledge to use AI tools.
\qs{The only risk you're getting when you're telling someone that you're using AI tool(s), is that he's going to request you to show a demo because they also want to use it.}{P26}
The only perceived \emph{risk} is the positive pressure of demonstrating tool capabilities to interested colleagues.

In our survey responses, geographic region again seemed to play a role in establishing a culture of sharing.
For the question about whether organizations that create open spaces for employees to discuss about AI tools increased the use of AI tools, 80\% (60/75) participants from Asia agreed, which is significantly higher than the other continents of Europe (57.8\%, 26/45), North America (61.9\%, 13/21), South America (54.5\%, 6/11).
Similarly, regarding whether participants were part of an organization that has strong culture of sharing insights, 79.5\% (62/78) participants from Asia agreed, again higher than Europe (73.6\%, 39/53), North America (66.0\%, 35/53), and South America (68.8\%, 11/16).
These findings indicated that culture seems to influence whether organizations and team members help facilitate an environment that is conducive for practitioners to share insights \cite{michailova2006national}.
Like IM2, collectivist cultures appeared more willing to create spaces for the flow of insights.


\subsubsection{OM2: Providing and promoting the tool:} \label{om2}
In addition to creating a culture a sharing, we found 11 participants \p{P5, P7, P8, P9, P11, P15, P17, P18, P19, P20, P25} revealing that their companies either paid for or showed willingness to cover the cost of subscriptions.  

Our interviewees explained that the decision to invest in paid AI tools is often driven by the potential for growth and the relatively minor financial burden of these expenses. 
P10 highlights that to many companies, the investment is not much.
\qs{what is the subscription fee for GPT for two months, something like 20 USD? From the company's perspective, it's nothing.}{P10}.
P10 further adds that their company usually pays for tool if the employee shows that it is essential to learn and grow. 
\qs{When I can convince them that [AI tool] is necessary, and it will help us grow a little bit further, the company will definitely will pay for it}{P10}
Interviewees whose organizations paid for their subscription argued that over the course of a month (i.e., 30 days), if developers could save at least a few hours of work, then the cost benefit of paying for a subscription would be positive. 

From our survey respondents we found that extra-large organizations were slightly most likely to pay for the use of tools 69.7\% (23/33). 
This intuitively makes sense as larger organizations would most likely have more resources at their disposal to cover the cost. 
We also found that the difference between the size of organizations was small in comparison with medium (63.3\%, 19/30) and large organizations (65.8\%, 25/38).
As one survey participant mentioned, organizational adoption can be improved when \textit{managers know what can they achieve easier now with lower cost, like if their adoption can increase the overall performance of company without increasing the number of staff.}
Our interviewees described that when their organization provides the tools, employees are empowered to use these tools as needed. 
\qs{Yeah, [company] pays also for the API usage and all of that. So not only copilot, but if you want to build like something internal, or just use the API for anything, they pay for it. ... and I know I can use it when I want to.}{P7}

In addition, organizations can enable employees to tailor models and versions for specific use cases and needs. 
There is also the ability to build add-ons and extensions from the tools. 
Prior to AI tools, this would be like an organization providing a custom Stack Overflow to its employees \cite{stackoverflow_teams}.
While previously difficult to achieve, this is now possible if organizations acquire subscriptions and API for its employees.
\qs{You can choose including several other models that you can pick from. 
We get either 1,000 or 500 instances of free conversation.}{P15}
Interviewees also mentioned that such enterprise versions of AI tools helps mitigate risks associated with data breaches and ensure a unified approach to tool usage. 
\qs{In the company, there are ready-made [tool(s)], and because you need to make extensive adjustments to your own account due to budget considerations, 
you will definitely use the company's unified planning.}{P18}

In addition to just providing AI tools, six interviewees \p{P2, P10, P15, P17, P20, P25} described how their organizations formally and informally promoted different tools.
Due to constant notifications, it is possible for practitioners to miss new tools introduced by their organization. 
\qs{We are bombarded with a lot of information every day, some people might turn off certain notifications. 
They might only find out about [new tool] after they noticed a colleague using something, and then they became aware of it.}{P17}
Since the Covid pandemic, many teams have been working in remote and/or hybrid settings, emphasizing the higher importance for coordination and communication \cite{de2022grounded}.
For organizations, the primary worry of not keeping up with industry practices is whether their developers are able to maintain a competitive edge.  
\qs{Our company encourages us to use those AI tool(s), because it heavily boosted our productivity. If we can't be familiar with those AI tool(s) and prompt engineering,  then it can be difficult for us in future to cope with the other developers in the world. 
}{P2}

Our interviewees discussed the role played by senior management who incorporated AI tools as part of the organization's broader strategic plan. 
\qs{The senior management really prioritize this matter. They want comprehensive intelligence, things like AI-driven initiatives. These are strategic directives coming from the top echelon.}{P15}
Similar to creating a culture of sharing, we find differences in promoting the tools depending on a participant's geographic location.
We found participants from Asia most often agreeing that their organization notified and motivated employees on using AI tools.
For OM2.3, 68.9\% (51/74) participants from Asia agreed, whereas  51\% (24/47) of participants from Europe and 55.3\% (21/38) from North America agreed.
Similarly, 62\% (44/71) participants from Asia, 45.2\% (19/42) from Europe and 55\% (18/33) from North America agreed with OM2.4.
 

\subsubsection{OM3: Providing guidance and training:} \label{om3}
Apart from sharing experience and providing and promoting AI tools, our participants also emphasized the importance of organizations providing guidance and training.
Of our interviewees, 16 \p{P2-P9, P13, P15, P17-P20, P23, P25} described the existence of some form of guidance related to using AI tools in their organizations.
Specifically, guidance refers to organizational policies and guidelines that govern what and how practitioners can use AI tools in their work.

Prior to the use of AI tools, organizations typically allocated some code of conduct for employees which included internal rules and policies. 
For example, an employee is forbidden to access customer data without authorization and forbidden to disclose company content without permission. 
This meant that an employee would not be able to create a set of company source code and upload it to GitHub or some other publicly facing file repository \cite{mergel2015open}.
Providing guidelines allows organizations to manage the concerns regarding responsible use of AI tool, especially pertaining to when using AI tools is appropriate and what tasks are not allowed.
Adding guidelines allows organizations to prescribe structure and expectations on handling sensitive data, which helps to prevent the leaking of company code and customer data.
\qs{In our company, it's not ok to release the code. There are also some very sensitive data involved, such as daily active user data those involve sensitive company information, which is also not permissible.}{P17}

Our interviews \p{P15, P19} indicated one approach to provide clear guidelines was dedicating resources to training practitioners.
Training covered best practices and techniques for prompting.
In P15's organization, senior management was heavily invested in pushing the entire company towards AI so training became mandatory. 
\qs{AI is essentially a strategy of the company, and requires everyone to get on board and take exams, earn credits, and then teaches you how to use these tool(s). Otherwise, it will be reported to the supervisor, and there will be some kind of reward and punishment measures in place.}{P15}
The organization not only provided training, but also required employees to demonstrate their knowledge through testing. 
Among the company's courses include discussions about \qs{certain sensitive topics that you might not be allowed to inquire [tool(s)] about.}{P15}
AI tools are rather new and still an emerging research area especially in areas such as prompt engineering. 
A number of active research is still exploring how to more effectively work with large language models, code generation \cite{chen2021evaluating, weyssow2023exploring, du2024evaluating}, and designing prompts \cite{reynolds2021prompt, brade2023promptify, liu2022design}. 
Unlike traditional software development tools such as Jira or GitHub or Visual Studio Code, which have remained stable over the years, AI tools are constantly undergoing change. 
P15 organization's training demonstrates the desire some organizations have to keep pace with the AI tool transformation.

Our survey respondents agreed that organizations providing training on prompting supports increased use and adoption of AI tools (79/123, 64.2\%).
In particular, surveyees from Asia (77.4\%, 41/53) strongly agreed that training supports increased use.
Contrastingly, participants from Europe (53.8\%, 14/26) and North America (58.6\%, 17/29) had much lower agreement. 
These numbers seem to align with earlier research on technology adoption that shows people from Asia were less resistant to trying out new technologies \cite{mckinsey2024, microsoft2024}.


In addition to providing guidelines and training, one direct strategy to tackle privacy exposure risk mentioned by interviewees is to develop a company portal that sanitizes all prompts. 
A company created portal allows employees to deal with customer data with an AI tool and lower the potential risk of data disclosure to a third party provider. 
\qs{You can also choose to use the internal resources of your company when you're dealing with customer data. You still have to use the interface of the company.}{P13}
Developing a company native portal is even more paramount for industries are under more security and privacy scrutiny.
For example, financial firms have to adhere to regulations such as PCI \cite{williams2022pci}.
\qs{Chatgpt, we use it as a policy that chatgpt is only allowed to use with level one data. So we have some kind of confidentiality level.}{P26}
To this end we found our interviewees acknowledging and making progress in sanitizing the prompts they send to external AI tools.

3 interviewees \p{P13, P15, P17} indicated that their companies have implemented structured mechanisms (i.e., sanitization) to ensure the safe and responsible use of AI tool(s) by their employees.
For instance, P15 highlighted
\qs{When I visit ChatGPT, I definitely have to go through the company's portal, and then on that end, there will be some verification and filtering happening. ... there may be a safety net where the company could help take care of it.}{P15}
Similarly, for P13, their sanitization tool first filters out sensitive information and can directly reject requests that may compromise safety in the input. There is a specific team in charge of the prompt filtration service and they use a combination of rule based and model based approach across multiple layers to remove compromising content. 
These measures prevent the sharing of confidential data, such as database passwords or proprietary code, while allowing for the use of AI for non-sensitive tasks, such as discussing code hierarchy or non-confidential snippets.
From our surveyees, 74/122 (60.7\%) of participants agreed that a tool to sanitize prompts helps protect company data.
Like providing training, surveyees who most strongly agreed about company providing tool to help sanitize prompt were based in Asia (71.4\%, 40/56), versus Europe (46.4\%, 13/28) and North America (54.2\%, 13/24).
Through providing sanitizations mechanisms and techniques, employees are encouraged to engage with AI tools while complying with security and confidentiality standards. 

\subsection{Organizational Challenges}
\subsubsection{OC1: Lack of culture of sharing:} \label{oc1}
While some organizations established a culture of sharing AI practices, our interviewees \p{P5, P8, P11, P13, P14, P22, P24, P25} also described how their organizations did not have a culture of sharing and collaboration. 
A lack of sharing is characterized by limited communication and interaction among colleagues about AI tool use and best practices.
Our interviewees highlighted that their discussion about AI tools was limited to individual use rather than summarizing lessons that could be disseminated company wide. 
Emerging in the post-pandemic work environment, previous literature showed that organizations working in remote and hybrid work environments, which are increasingly common, has to watch for coordination, collaboration, and communication challenges \cite{de2022grounded}.
Coordinating and communicating about software development can be challenging, without even considering \textbf{``what are the tools that my colleagues are using?"}
We found a frequent narrative was that practitioners are either unaware of their colleagues' use of tools or perceived the use of these tools as irrelevant to their field of work.
In particular 77.4\% (24/31) of survey participants agreed that everyone does different work and they do not share a similar project context as the reason for not communicating about AI best practices. 
68.9\% (20/29) of survey participants believed that there is no culture of sharing with their team or organization. 
For example, \qs{I feel that at least in our company, this kind of atmosphere hasn't come about, or it's how we view ourselves in this context.}{P11}

We found participants who have access to group messaging, but there is little engagement from teams that result in productive discussions about the tools.
Referring to their company group chat, \qs{I don't think so. No [sharing], not in my team.}{P8}
This results in small, unorganized discussion of tips and best practices rather than meaningful discourse that may result in a coordinated organizational wide strategy. 
P25 exemplifies this problem about how a team does not know what tool(s) are being used in nearby teams.
\qs{I think the other two teams don't even know about it because the communication it's not really good.}{P25}

Another limiting factor is that participants may be dissuaded when they perceive an organization is against sharing of experiences. 
\qs{Absolutely not. There's no culture there. I think like, I know that some people on my team... they don't know that it's out there... There's no culture of sharing or anything like that.}{P5}
\qs{No, there are very few [sharing activities]. 
It might be that the whole company hasn't really promoted or initiated such activities.  
Someone at my level, a middle-tier leader, will relatively follow the company's policies in doing things, making plans.}{P14}
In P14's case, even though they are a technical lead with dozens of employees working under their supervision, they still need to follow the company's lead. 
Therefore, if an organization demonstrates limited interest in promoting a culture of knowledge sharing, the effect seems to cascade down. 


\subsubsection{OC2: Company does not cover the cost of the tool:}
A significant concern that emerged from our interviewees and surveyees was the cost of the premium version of the AI tool(s). 
8 interviewees  (P5, P6, P11, P12, P13, P14, P22, P26) indicated that they bore the expense of these tool(s) out of their own pockets to have access to better features. 
120/201 (59.7\%) survey participants reported that their AI usage is challenged due to the high prices associated with subscriptions and paying for it. 
In particular, we found a discrepancy between geographic location regarding the payment of tools perceived as a problem.
For survey participants based in Asia, 76\% (57/75) agreed that their usage of AI tools is limited by the high cost of AI tools and paying.
There was a significant gap with the participants from Europe (52.9\%, 27/51) and North America (37.8\%, 17/45).
For instance, P16's organization is based in Asia and has a large number of employees, but the organization was unable to cover the cost of the tool due to payment. \qs{The company is relatively poor. We don't have this paid version!}{P16} 
Many of the largest AI tool providers are based in North America so it seems reasonable for organizations based there to establish partnerships with these providers \cite{hbr2023}.

Employees may purchase a subscription as a trial run and try it out for a period of time, but the high cost of subscriptions can lead people to just revert back to a standard free version. 
\qs{During the initial stage of trying it out, I did purchase the pro version. But, the company didn't buy this kind of license. This is a reflection of the company's real situation.}{P14}
Not having access to the full paid version could have negative repercussions on the benefits a practitioners could realize given that paid tools usually provide bigger prompt limits, more recent knowledge base, and ability to more frequently query \cite{diaz_chatgpt_nodate}. 
The paid version often offers better features which can help accelerate the development work.

Related to the paid versus free versions, is the number of API calls granted per unit of time. 
As of this time of writing, a plus version of ChatGPT allows up to forty prompts per three hours.
A developer can quickly reach the upper bound if they are conducting a lot of experimentation.
73/142 (51.4\%) of survey respondents reported that the number of API calls allowed in a time interval is limiting their usage of AI tools.
In contrast, an organization such as P15's employer grants developers up to 500 or 1000 prompts per day. 
We found among the survey respondents that those participants who had less than 1 year experience more commonly reported the API problem with 75\% (12/16) versus 52.5\% (21/40) for participants with 5-10 years experience and 43.3\% (13/30) for participants with 10+ years experience. 
One cause for this discrepancy could be that juniors have less experience with software development so they may require more trial and error leading to higher API use. 
Seniors, in contrast, have a wealth of experience and require less trial and error prompting so they are more likely to reach the solution using less prompts.


\subsubsection{OC3: Lack of company guideline:} \label{oc3}
In addition to some organizations not covering the cost of AI tools, we also found 9 interviewees \p{P10, P11, P12, P14, P16, P21, P22, P24, P26} who indicated that there is a lack of formal training and guidelines in their company, on the effective use of AI tools.
Prior to AI tools, organizations implemented restrictions preventing unauthorized sharing of sensitive information on platforms like GitHub, but no such guidelines have been prescribed to address the emerging challenges of AI tools.
\qs{In the early days it was mentioned that it was strictly forbidden to post anything on [Github]. And it should be in the handbook. The company have not made any specific and clear policies targeting AI tool(s).}{P14}
These interviewees complained that their organizations adopted a laissez-faire attitude towards governing the use of AI tools.


Sometimes not publishing any guidelines is a purposeful decision, because a company may be unsure about how these tools handle data.
\qs{We have a monthly meeting where me and my colleagues gave a presentation about ChatGPT how to use it in the work. 
However, there is not an concrete example, this is what you're able to do. 
I think they one side is that they don't know yet.}{P21}
P21 further adds, 
\qs{We had a training about security, about what to do and how to not get your information leaked, the basic things. But for AI, it's still not that explicit.}{P21}
One positive and one negative consequence is caused by this laissez-faire attitude. 
The positive is that employees may choose to use AI tools how they see fit, the negative is that employees are operating in a grey area of company policy. 
36.2\% (38/105) of participants in small organizations and 40.5\% (15/37) of participants in medium organizations agreed that their use of AI tools was limited due to company not providing any usage guidelines,
in contrast to 29.6\% (8/27) and 21.9\% (7/32) of participants from large and extra-large organizations. 
Larger organizations tend to have more structure in place compared to smaller organizations who have flatter organizational structure \cite{cummings1976organization}.
For example, during the push for European organizations to comply with the General Data Protection Regulation \cite{gdpr}, one of the most stringent privacy regulations in the world, smaller and medium organizations were found to be struggling to comply more than their larger counterparts \cite{sirur_are_2018}. 
For guidelines surrounding AI tools, smaller organizations again may not have the necessary resources to form dedicated teams to analyze data collection and processing conduct from AI tool companies.

Aside from the lack of guidelines, we found 20 participants \p{P5-P14, P16-P18, P20-P26} indicating that there was a lack of formal training from their company on the effective use of AI tools.
Participant P25 noted that AI tool(s) were introduced to their company's Slack with no training or guidance. \qs{It was dropped just like that.}{P25} 
Employees in the company were left to independently navigate and explore the tool's functionalities without any organized support from the company.
Our interviewees' perception was that training could enhance developers' proficiency in crafting prompts and leveraging AI tool(s)' capabilities.
\qs{if there's a training, which is able to efficiently help you like software developers, specifically, and how to prompt, I think it will be very helpful.}{P21}
49/197 (24.9\%) survey participants felt that their use of AI tools was challenged from a lack of training provided by their company.
This ratio is not considerably high, but we note that most AI assistance tool(s) for software are relatively easy to get a conversation started \cite{borsci2022chatbot, jo2023decoding}. 
Training to use the basic functionality of these tools is most likely unnecessary.
However, as described in Subsection \ref{om3}, training can help employees become more proficient in prompting. 
Like Subsection \ref{om3}, we found noticeable differences in survey demographics.
Survey participants from Asia (32.9\%, 22/67) agreed more often than those from North America (17.3\%, 9/52) and Europe (13\%, 7/54) that company not providing training limited use of AI tools.

11 interviewees \p{P1, P4, P10, P13, P14, P17, P20-P23, P26} voiced concern regarding the exposure of company data privacy as a challenge that limits their adoption of AI tool(s). 
In recent years, a growing push from legislators around the world was enacting stringent privacy regulations in an effort to more effectively protect user privacy. 
Among the complex to navigate regulations include the GDPR \cite{gdpr}, California Consumer Privacy Act \cite{ccpa}, Personal Data Protection Act \cite{pdpa}, Personal Information Protection Law \cite{pipl}, and Consumer Privacy Protection Act \cite{cppa}.
Famous cases of privacy infringements resulting in loss of user data also costed company reputation and financial penalties \cite{cso2023}.
It was reasonable for interviewees to express worry about the potential privacy consequences that may arise from the use of AI tools.
There is also some concern that policy makers are slow to react to technology changes. 
\qs{For those who are against such technology, for lawmakers, it’s actually because we often know that legislation usually lags behind. 
}{P14}

Previous literature has covered concerns regarding the integration of AI tools like Copilot \cite{jaworski2023study, zhang2023practices} and the security worries from their use. 
We found one primary concern from participants is that AI tools are recording user prompts and then using the prompt contents as further training data.
\qs{Data security is really concerning part for us, and that's why we can't depend on ChatGPT and we can't use in our IDE. What will be the use cases of the consumed data and whats the guarantee they will not leak it somehow and sell it to somewhere. Are they consuming my credentials? Are they consuming my personal data?
}{P4}
Moreover, many users of these AI tools rarely understand terms of service and privacy policies even when they are available. 
This is a notorious problem among research regarding how to ensure end users understand privacy risks \cite{tang2021defining}.

We found in our survey responses that there was a noticeable difference in attitude between participants from larger organizations in comparison to those from smaller organization. 
68.4\% (26/38) and 59.4\% (19/32) from extra-large and large organizations agreed that their use of AI was limited by privacy versus 35.9\% (37/103) and 39.4\% (11/28) from smaller and medium organizations. 
The deep-rooted concern with privacy is largely attributed to customer data.
Consequently, participants expressed hesitance to use AI tool(s) for tasks that might involve any customer confidential data. 
\qs{Data from conversations. We will try not to use real ones. For example, we also have e-commerce platforms. The data from dialogues between customer service and users, we will not use those. 
Especially when it involves buyers' addresses, phone numbers, and such.}{P17}

Given the significant fines that have been handed out to companies for violation of privacy regulations, \cite{cso2023} it seems reasonable for employees from large organizations to be extra considerate about user data.
The most significant fines against organizations have largely all against large organizations with significant financial resources.
Based on the GDPR fines, smaller organizations have for the most part avoided excessive scrutiny.
Practitioners in smaller organizations may not have organizational guidelines governing AI tool terms of use but are less worried about privacy because \textbf{they may not survive.}
\qs{If I am a large corporate, I wouldn't want user data to leak. For a small company, surviving until the day we receive a legal notice is one thing, but if you're a large company and let's say you have a major user data breach, that's a huge problem.}{P26}
Ultimately, privacy regulations that are passed by jurisdictions have strong implications for how much companies fear data breaches from using AI tools. 
Similar to the Cambridge Analytica case \cite{nyt2018}, one privacy breach can have irreparable damage to reputation and cause negative financial consequences.

\begin{table}[]
    \centering
    \small
    \caption{Relationships between factors from interviewees and survey voting results (from 5-step Likert scale)}
    \begin{tabular}{p{1cm}p{3.5cm}p{5.9cm}p{2cm}} \toprule
       ID  & Relationships & Survey Prompt & \par Top-3 \#Votes \strut (Ratios) \\ \midrule
        \multirow{2}{*}{\parbox{1cm}{IM2 \\ \& OM1 \\  \& IC4 \\ \& OC1}} &   
        \multirow{2}{*}{\parbox{3.5cm}{
        Culture of sharing AI knowledge vs. Negative judgment and lack of support}} 
        & If organization builds a culture of sharing AI practices would it increase your use of AI tool(s)? & \parbox{2cm}{173/223 (77.6\%)} \\ 
        & & If organization builds a culture of sharing AI practices would it reduce the potential for negative judgment by others? & \parbox{2cm}{158/216 (73.1\%)} \\
        \midrule

        OM2  \& OC2 & 
        Providing and promoting vs. Not paying the cost & If organization starts paying for employee use of AI tool(s) would it increase employee's use of AI tool(s) & \parbox{2cm}{193/226 (85.4\%)} \\\midrule

        \multirow{2}{*}{\parbox{1cm}{OM3 \\ \& IC3 \\ \& OC3}}  & \multirow{2}{*}{\parbox{3.5cm}{
        Providing guidance and training vs. Lack of prompting skills and usage guidelines
        }} & If organization implements a policy on the rules, restrictions, and accepted terms for using AI tool(s) would it increase your use of AI tool(s)? & 131/205 (63.9\%) \\
        & & If organization provides guidance and training for AI tool(s) would it increase your use of AI tool(s)? & 168/231 (72.7\%) \\
        & &  If organization provides tool(s) and techniques for prompt sanitization would it reduce your concern about data privacy leaks? & \parbox{2cm}{176/227 (77.5\%)} \\

     \bottomrule
    \end{tabular}
    \label{tab:relationships}
\end{table}

\section{Push-Pull Relationships between the Motives and Challenges}
This section describes relationships that we observed between some of the motives and challenges that have significant impact on the use and adoption of AI tools.  
These relationships act in a push-pull manner, whereby 
motives and challenges are interconnected with each other. 
Motives \emph{pull} practitioners and organizations to increase use and adoption of AI tools, whereas challenges act in the opposite way and \emph{push} practitioners and organizations away from adopting and using AI tools.
For example, \emph{providing and promoting the tool} motivates the use and adoption of AI tools, but in contrast, \emph{company does not cover the cost of the tool} inhibits use and adoption.
We derived these relationships because they help indicate the factors that software organizations can consider to improve their use of AI tools.

\subsection{Culture of sharing AI knowledge vs. Negative judgment and lack of support:}
One of the most important relationships identified in our research was \textbf{establishing a culture of AI experience and knowledge sharing helps increase AI tool use and adoption, but a lack of support and potential judgments impede adoption and use.} 
This relationship intersects all four categories identified in our theory from individual motive, and organizational motive, to individual challenge, and organizational challenge. 

From an individual perspective, we found that participants often had a natural inclination to share their experiences and best practices from their use of AI tools.
The willingness to share knowledge not only fosters a collaborative environment but also informs colleagues who may not be aware of the productivity boost. 
\qs{People give ideas about the best way to prompt ChatGPT.
I think that was usually somebody will share cheat sheets. This was something that really helped them like over time. And people kind of like, keyed into that and started to use that.}{P20}
This type of individual initiative was also on display from practitioners who actively encouraged others in their organization to start using AI tools. 
\qs{[Employees] encourage other people to use it. But it doesn't come from an organizational level where it doesn't come from the top down to say we should be using these to make us more efficient.}{P9}

However, this individual motive is limited by a counter factor, primarily fuelled by a fear of judgment from other team members. 
This fear stems from a perception that reliance on AI tools may be seen as a lack of capability or laziness.
\qs{Because you're relying on it so much that you're not using your brain, which is that's a problem.}{P6}
The criticism is typically directed towards junior developers who leverage AI tool(s) to help them learn, search, and debug code. 
\qs{I've seen that passing judgment. 
Some of them is still passing judgments when using the ChatGPT.}{P4}

In comparison, 73.1\% (158/216) of survey participants selected the top 3 levels in a 5 level Likert scale indicating that if an organization builds a culture of sharing AI practices, it would reduce the potential for negative judgment from using AI tools.
Some of our participant organizations facilitated a culture of sharing, where team members were encouraged to discuss their experiences and share tips. 
This sharing of knowledge is often conducted through communication channels such as Microsoft Teams or Slack, which act as avenues for knowledge transfer, as described by P7 and P21 in Section \ref{om1}.
78.3\% (173/221) of survey respondents agreed that an organization that builds a culture of sharing AI practices would increase their use of AI tools.
Meanwhile, the chief reasoning for a lack of sharing was that employees conducted different work and organizations did not create an atmosphere conducive for sharing with other team members.
\qs{It's because perhaps everyone has different habits, and so each individual faces different problems, including even the fact that everyone is responsible for different modules.}{P17}
Similarly, for P22, colleagues have few discussions about AI tool use.
\qs{None of the people like chatting.}{P22}
In such environments, conversations about best practices, tips, and experiences are limited, and there are less opportunities for disseminating AI tool related knowledge within the organization.


\subsection{Providing and promoting vs. Not paying the cost:}
Another relationship identified in our work was \textbf{providing and promoting AI tools within an organization helps increase AI tool use and adoption, but adoption and use is impeded when organizations do not pay for the cost of AI tools}
We found participants agreeing that organizations covered the cost of AI tools was a motivating factor for employees to use them. 
When companies do not cover the cost, fewer people use the tools because the cost becomes a hurdle. 
Depending on geographic location and available resources, the cost may be significant for a practitioner, especially as the number of AI tools increases every time a major AI company releases a new model or tool. 
193/226 (85.4\%) of our survey respondents indicated that employee use of AI tools would increase if an organization starts paying for them.

One trial and error approach seemed to work for some participants.
An organization may not initially pay for a tool, \textbf{if an employee can show that the tool is necessary for their job and boosts productivity, the company might then agree to cover the cost.}
\qs{If I can prove that without chatGPT or without the paid version, I cannot work or it will help a lot for me, as well as the company to work along the way. If I'm using the paid version of chatGPT, the company will definitely pay for it.}{P10}

\subsection{Providing guidance and training vs. Lack of prompting skills and usage guidelines:}
The final use and adoption relationship we found is \textbf{providing guidance and training for using AI tools help increase AI tool use and adoption, but adoption and use is impeded when individuals lack prompting skills and organizations do not provide AI tool usage guidelines}.
As described earlier in OM2 in Section \ref{om3}, when organizations invest in training for developers about AI tool(s) and prescribe clear guidelines, it clarifies the limits of what employees can and cannot do. 
This can help reduce uncertainty where a developer may not know if their actions are legally permitted by their organization. 
P21 provides an example of how he believes training would help. \qs{I think if there is training, which is able to efficiently help you like software developers specifically, and how to prompt, I think it will be very helpful.}{P21}

AI is a rapidly changing field with a plethora of active research. 
Sometimes within the span of weeks, major updates are made to popular AI tools or a completely new AI model is released altogether. 
Consequently, it can be difficult for practitioners to keep up with the pace of AI change if they are not actively paying attention to all the changes \cite{yahoo2024}.
\qs{The majority of my fellow colleagues, including myself, might not know in what situations AI tool(s) can play a certain role, and also, with the current large models at the front line, their speed of upgrading and iteration is very fast...}{P15}
Should an organization provide prompt engineering and other training, it would help guide practitioners to understand how to best use the available tools and which software development task work best for each tool or model.
72.7\% (168/231) of survey participants agreed that if an organization provides guidance and training for AI tool(s) it would increase their use of AI tool(s).


In contrast, our work found challenges pertaining to when organizations failed to provide clear guidelines on the use of AI tool(s), as discussed in detail in Subsection \ref{oc3}.
When there is a lack of clear guidelines, many practitioners suffered from a  fear stemming from an undesire to inadvertently violate company policies, which could lead to negative consequences for both the individual and organization. 
\qs{I think we should be more careful about compliance this and things like that, 
I think sometimes in software dev, you kind of forget, because you're more lenient on looking for things online.}{P23}
Correspondingly, 63.9\% (131/205) of survey respondents agreed that their use and adoption of AI tools would increase if an organization implements a policy on the rules, restrictions, and accepted terms for using AI tools.

Additionally, we found that when organizations implement measures to sanitize inputs and outputs of AI tool(s) described in section \ref{om3}, developers feel more comfortable. 
These sanitization processes are particularly important for practitioners to be assured that sensitive information, such as API keys, user data, or proprietary code, is not inadvertently captured and misused by these tool(s). 
\qs{We also maintain partnerships with organizations like OpenAI... Could involve filtering out safety or safety issue-related inputs and outputs.}{P13}
The absence of such measures leaves developers wary about using AI tool(s) with sensitive or confidential code. 
176/227 (77.5\%) of survey participants indicated that if an organization provides tool(s) and techniques for prompt sanitization it would reduce their concern about data privacy leaks.

The absence of such measures leaves developers wary about using AI tool(s) with sensitive or confidential code. 
\qs{I never use names or brands, especially in terms of my company, I never use any of the company names. I usually obscure even references to environment variables.
It's very important that there's no reference to any real person.}{P22}
This fear often leads them to alter code or create a hypothetical situation in their prompts to avoid the risk of exposing sensitive data.
\qs{I'll always have a sanity check is this. Will this be okay to put in ChatGPT? I don't just copy-paste and put it in there. I sometimes do like pseudocode.}{P21}
Therefore, our participants suggest organizations implement a combination of clear policies governing tool use and providing means of filtering sensitive data directly inside prompts. Otherwise, the risk of privacy exposures can be costly \cite{gdpr_amazon, gdpr_meta1, gdpr_meta2, gdpr_fines}. 


\section{Discussion}
Previous research studying AI tool(s) in software development has often focused on the tool(s) themselves and the usability of those tool(s) for software tasks \cite{liang2024large, StackOverflowSurvey2023, peng2023impact, gitlab2023devsecops}.
These prior works found that developers reportedly gain immense boosts in productivity when they leveraged AI assistance to help with development \cite{peng2023impact, liang2024large, gitlab2023devsecops}.
As a result, AI tool(s) such as ChatGPT experienced monumental user growth, where in less than 1 year, it successfully attracted 100 million weekly users \cite{porter2023chatgpt}.

One previous study attempted to provide more contextualization regarding the adoption of generative AI \cite{russo2023navigating}. 
Russo developed a preliminary theory for the adoption of generative AI, however, 3/7 of his hypotheses were rejected upon analysis \cite{russo2023navigating}.
In particular, he suggested that organizations play a limited role in impacting developer adoption of AI tools, that more future empirical work is needed to investigate this fast-changing landscape. 
Therefore, we still lacked clarity on the impact of organizations and individuals gravitating toward or away from adopting these tools. 

In our work, we identified challenges and motives that push and pull practitioners into using AI tool(s).
Unlike individual motives, which may have a visible impact, our study also uncovered the importance that organizational culture serves in the adoption and increased use of tool(s).
Our findings on the impact of organizations contradict with previous literature \cite{russo2023navigating}, however, we surmise that this may be a result of our emphasis in our study about organizational role in the use and adoption of AI tool(s).
Moreover, it is possible that aspects such as guidelines and data privacy are becoming more important due to ongoing lawsuits and regulatory concerns.

In our study, we aimed to fulfill this gap through a Socio-Technical Grounded Theory approach, which included 26 interviews with practitioners and a survey with 395 respondents regarding their use and adoption of AI tool(s) for software development. 
Our findings presented a list of 12 factors including 7 challenges that influence the use and adoption of AI tool(s).
We also presented 3 relationships between factors, which detailed how increasing motives for use and adoption of AI tool(s) may help reduce its corresponding challenges. 

In the following subsections, we discuss how to increase the use and adoption of AI tools in organizations and the importance of privacy concerns and ethics amidst the use and adoption of AI tools.
Finally, we propose directions for future research. 

\subsection{Increasing adoption of AI tool(s) in organizations}
The primary finding in our study is what are the factors that are motivating and challenging the adoption and usage of AI tool(s) for software development.
Previous works have shown that developers who use tool(s) like Copilot and ChatGPT should reap significant productivity improvements and become more effective in their work \cite{kalliamvakou_research_2022}.
While there is debate on exact extent of improvement, the current literature and confirmed in our study supports the notion that AI tool(s) boost practitioner productivity.

However, despite these benefits, our study unveiled that not all companies or developers are embracing AI tool(s) to their fullest potential. 
Several limiting factors were identified, affecting both the extent of adoption and usage. 
At an individual level, knowledge gaps and uncertainty about the optimal use of AI tool(s) can hinder their adoption.
We found that AI tool adoption in organizations is not trivial as simply creating a ChatGPT account and using it to generate software code. 

Previous works have stated productivity boosts \cite{peng2023impact, liang2024large}, materializing from auto completion, finding code solutions and edge cases are motivations for developer use of AI tool(s).
Whereas motivations for not using AI tool(s) include lack of useful or relevant output, particularly involving functional and non-functional requirements \cite{liang2024large}.

One organizational challenge that has profound impact on software professionals is simply organizations not covering the cost of AI tool(s) such as ChatGPT or others.
Although the cost of \$20 USD may seem trivial to resource rich organizations, numerous participants in our study indicated that their organizations are not covering this cost.
For participants who are in less resource rich situations, the price of ``Plus" subscriptions may be cost prohibitive.
As one surveyee explained with respect to what practices their organization should follow to improve the adoption, \emph{``Offer cost reductions for AI tool(s) and even assist employees in covering the expenses."}

Even for organizations who provide AI tool(s) directly to users the choice of subscription could also have negative ramifications.
For example, P25's organization provides a company subscription to ChatGPT integrated into Slack, but it is the lower 3.5 version. 
Consequently, P25's motivation to use the AI tool was dampened due to poor generated results. 
\emph{``But it just that version, it just repeats things that I asked and answers me back with the same words, like scrambled, I didn't like it. It didn't get to the point for the documentation for me. It's just babble."}

From an organizational perspective, the culture surrounding knowledge sharing plays a pivotal role. 
In environments where sharing and collaboration are encouraged, knowledge of AI tool(s) and their potential benefits disseminates more rapidly, fostering a more widespread and effective adoption.
When employees share insights with others in the company or team, it can significantly improve the productivity of colleagues because some employees are not aware of the best practices.
In P16's organization, colleagues share insights about how to use tool(s) to increase productivity.
As an example, one colleague demonstrated how they were able to use AI tool(s) to develop a useful web scraper in 15 minutes, the previous human manual effort would have been a whole day. 

In contrast, practitioners should be cautious of potential judgment and lack of culture of sharing manifesting in their organization as they may limit the gains an organization could realize. 
As one surveyee respondent added, \emph{``Driving AI adoption from the top down is the only way to get people 'out of the shadows'. And the only way to do that is for leaders to spend the time and curiosity to understand limitations and opportunities from AI, and encourage that sharing in the org."}
Organizations should avoid creating environments where employees are negatively judged and rated depending on their usage of AI.
One survey participant explains that the reason they do not share their experience of using AI tool(s) with other colleagues is a result of, \emph{``Just they might judge my performance if I say I do so}."
Our findings emphasize the importance of addressing both individual and organizational challenges to maximize the benefits of AI tool(s) in software development. 
Software professionals can use our theory to identify areas to watch out and practices to adapt to increase motives for high use of AI tool(s).

\subsection{Privacy concerns and ethics}
Previous literature focusing on the ethics \cite{pant2024ethics} of AI practitioners have shown that practitioners are aware of the risks of privacy. 
In fact, \emph{``Privacy protection and security"} was the concept that majority of AI practitioners in the study were aware about \cite{pant2024ethics}.
A separate multiple case study on ethical concerns from implementing big data and AI found that privacy was the only ethical principle that was prevalent in all the case studies \cite{ryan2021research}.
Despite the recognition of need for ethical considerations by those who \emph{build} AI products, our findings indicate that organizations and practitioners are nonetheless still significantly concerned about privacy.

A large number of our respondents emphasized that they are limited in their adoption and subsequent use of various AI tool(s) because they are uncomfortable sharing company data in the form of code or documentation, as well as feeding customer data directly to a tool.
As one survey respondent explained, \emph{``Uncertain data privacy both internally and externally with AI providers/tool(s)"}.
A previous survey on usability of AI programming assistants found that 54\% of participants gave up on using AI assistants because code generation tool(s) did not meet functional or non-functional requirements, which includes security and other requirements \cite{liang2024large}.
However, our findings specifically highlight the anxiety surrounding privacy.

Not only are respondents concerned about prompting itself, but also worried about the unethical sourcing of data.
As one surveyee described, \emph{``Major privacy problems. Also data sourcing, most data collected by AI has been done unethically without consent."}
Similarly, P8 also pointed out this uncertainty, \emph{``So just always want to make sure that whatever you whatever I input doesn't contain any kind anything that is tied to the company... It's an internal policy. Because I think that data is also feedback to like train the model to be more efficient."}

As another survey respondent elaborated, more to this problem about data collection issue is that \emph{``There are few LLM-based tool(s) which have clear privacy manifests and have strict rules to a data collection, that is the biggest show-stopper. There is little to no knowledge on how models were trained as well, which leads to a question on how legal is it to use output of such LL models."}
From the perspective of software professionals and their organizations, it would help dissuade their concerns if AI tool developers began clearly outlining what they are collecting and how the security will be handled. 
\qs{They should outline clearly what they are consuming and how they're consuming and what will be the security of this the consume data.}{P4}

However, the privacy discussion is ongoing, where the development of stringent regulations is being discussed in various jurisdictions around the world \cite{noauthor_artificial_2024, warren_legalweek_2024, yang_four_2024} amidst calls for more regulation  \cite{open_letter,diaz2023connecting,hacker2023regulating, finocchiaro2023regulation,pascual_geoffrey_2023}. 
Advocates call for auditing and regulation of AI \cite{diaz2023connecting, open_letter}, with some industry giants even calling for an immediate pause to AI experiments \cite{open_letter}.
In contrary, there are also some debate about balance to strike between constraints that may stifle innovation.
\qs{If you impose constraints too early and too strictly, you end up limiting what it could achieve in the future, where it could go. We can see that LLMs have already made some sort of breakthrough developments in certain areas. [With stringent regulation] they might find themselves in an awkward competitive spot, and may not be willing to invest in researching and studying this technology.}{P14}

However, our findings still provide some suggestions that software organizations and practitioners may consider to assist their adoption and mitigate some concerns.
Software organizations may alleviate some uncertainty internal to organizations by establishing clear guidelines governing the use of AI tool(s).
These guidelines may include aspects such as names of the tool(s) permitted by the organization.
Moreover, the types of prompts and the types of data permitted as input should also help alleviate practitioner confusion.

Software organizations may also opt to develop or externally acquire tool(s) that plugin into the AI tool(s) that support sanitizing or anonymizing practitioner prompts so that confidential or sensitive information is filtered before sending.
For example P15's organization established a sanitization tool that filters a prompt of sensitive content before sending it off to a third party tool. 
This sanitization helps abstract software professionals from checking themselves if their prompt satisfies organizational privacy requirements.

\subsection{Directions for Future Research}
\begin{itemize}
    \item \textbf{Prompt sanitization and privacy safeguards:} Our research highlights the importance that many practitioners place on effectively protecting company and user privacy. 
Privacy and security of prompts when using AI tool(s) is hugely critical for more widespread adoption and use of these tool(s).
Further research could investigate leveraging security and privacy mechanisms to protect confidential data as built-in plugins when using AI tool(s).

    \item \textbf{Organizational aspects:} We also found that organization play a significant role influencing the use and adoption of AI tool(s). Whether it is the payment for AI tool subscriptions or facilitating a culture of AI practice sharing, organizational culture impacts the decisions that practitioners make about AI tool(s). More empirical research into organizational aspects such as culture and size are warranted to better understand how these factors prevail and their specific impacts on practitioners.
More importantly, more research into this area could develop strategies into mitigating the challenges with adoption and usage that we identified in this work.

    \item \textbf{Education:} The participants of our study expressed how training and understanding prompting is crucial to achieving the most benefits out of the tool(s).
While some organizations provide varying levels of guidance and basic training for producing more effective prompts, more research is needed to create more structured training.
Future research could study the areas and content that should be part of manuals and training material to help practitioners become effective users of AI tool(s).

\end{itemize}

\subsection{Threats to Validity}
Despite our best efforts in mitigating threats to our research, there are still some limitations. 
We use the total reliability framework from Roller \cite{roller_applied_2015} and the aspects of credibility, analyzability, transparency, and usefulness to assess the threats to validity. 

\emph{Credibility} refers to the accuracy and truthfulness of the data. It is about ensuring that our research accurately represent the data collected and the perspectives of our study participants. 
For our research, our selection of interview participants is limited by our convenience sampling, which means reaching out to practitioners from our personal contacts. 
However, we mitigated this threat by ensuring that each participant matches our selection criteria, which is aimed towards answering our research goal.
We ensured that each participant was a practitioner who uses AI tool(s) in their software development work.
In addition, we relied on a mixed methods approach where we validated the findings from our interviews with a survey towards software professionals.

\emph{Analyzability} refers to the ability to draw meaningful conclusions from the data.
For our study, we relied on the transcription tool Otter.ai \cite{otter} to help transcribe each interview. 
We also relied on SurveyMonkey to conduct our questionnaire survey \cite{surveymonkey}.
Three co-authors followed the steps of STGT to collect and analyze the data, including conducting open coding, memoing and constant comparisons until reaching saturation at which point the theory was developed.

\emph{Transparency} refers to the the clear and completeness of documenting of the research process.
For our study, we provided thick descriptions of our entire research methodology.
We also provided examples and used quotes wherever possible. 
Our provided as much details as possible to show the relationships between the themes in the findings. 
In addition, we release a replication package containing the interview questions, code book, and survey questions.
Unfortunately, we are bound by research ethics obtained from our institution and unable to release raw transcripts nor answers from interview and survey participants.

\emph{Usefulness} refers to the utility of our research and whether the findings have value to the relevant audience. 
AI assistance tool(s) for software is currently undergoing a rapid pace of change, which is challenging for our research. 
However, many of our findings transcends cultural aspects that are not easily \emph{``solved"} by an update in an AI tool, but otherwise require further study to mitigate any of the challenges limiting adoption of AI tool(s).
This is why we conducted both interviews with diverse software professionals and also conducted a validation round of questionnaire surveys with over 395 professionals from around the world. 
Although we do not assume that every software organization will encounter the same challenges and motives identified in our research, we expect professionals and organizations that share characteristics as those in our study to experience some similar benefits and limitations.

\section{Conclusion}
Practitioners and organizations are increasingly relying on AI assistance tool(s) to help with software development, recognizing the importance these tool(s) play in the present and future.
We followed the steps of Socio-Technical Grounded Theory to guide our research through 26 interviews with industrial practitioners and theory validation with 395 survey respondents.
Our findings show that practitioners are motivated to adopt and use AI tool(s) based on 2 individual motives and 3 organizational motives, but their use is also challenged by 4 individual  and 3 organization factors.
9 of these challenges and motives are linked across 3 push-pull relationships.  
Given the benefits that practitioners may acquire from using AI tool(s) in software development tasks, we highlight the challenges which may limit increased adoption and use.
We also emphasize the importance for organizations to create guidelines to help clarify potential privacy concerns and ethics from practitioners.

\bibliographystyle{ACM-Reference-Format}
\bibliography{./main}


\end{document}